\documentclass[fleqn,usenatbib]{mnras}

\usepackage{newtxtext,newtxmath}
\usepackage{multirow}
\usepackage{cleveref}
\usepackage{tensor}
\usepackage{amsmath}

\usepackage[T1]{fontenc}

\DeclareRobustCommand{\VAN}[3]{#2}
\let\VANthebibliography\thebibliography
\def\thebibliography{\DeclareRobustCommand{\VAN}[3]{##3}\VANthebibliography}

\usepackage{graphicx}	
\usepackage{amsmath}	

\title[MGEs for strong lensing]{Unveiling Lens Light Complexity with A Novel Multi-Gaussian Expansion Approach for Strong Gravitational Lensing}

\author[Q. He et al.]{Qiuhan He$^{1}$\thanks{E-mail: qiuhan.he@durham.ac.uk},
James W. Nightingale$^{1,2,3}$,
Aris Amvrosiadis$^{1}$,
Andrew Robertson$^{4}$,
Shaun Cole$^{1}$,
\newauthor
Carlos S. Frenk$^{1}$,
Richard Massey$^{1, 2}$,
Ran Li$^{5, 6}$,
Xiaoyue Cao$^{5, 6}$,
Samuel C. Lange$^{1}$,
João Paulo C. França$^{7}$
\\
$^{1}$Institute for Computational Cosmology, Department of Physics, Durham University, South Road, Durham DH1 3LE, UK \\
$^{2}$Centre for Extragalactic Astronomy, Department of Physics, Durham University, South Rd, Durham, DH1 3LE, UK \\
$^{3}$School of Mathematics, Statistics and Physics, Newcastle University, Newcastle upon Tyne, NE1 7RU, UK \\
$^{4}$Jet Propulsion Laboratory, California Institute of Technology, 4800 Oak Grove Drive, Pasadena, CA 91109, USA \\
$^{5}$National Astronomical Observatories, Chinese Academy of Sciences, 20A Datun Road, Chaoyang District, Beijing 100012, China \\
$^{6}$School of Astronomy and Space Science, University of Chinese Academy of Sciences, Beijing 100049, China \\
$^{7}$Centro Brasileiro de Pesquisas Físicas, Rua Dr. Xavier Sigaud 150, Rio de Janeiro, 22290-180, Brazil \\
}

\date{Accepted XXX. Received YYY; in original form ZZZ}

\pubyear{2015}

\begin{document}
\label{firstpage}
\pagerange{\pageref{firstpage}--\pageref{lastpage}}
\maketitle

\begin{abstract}
In a strong gravitational lensing system, the distorted light from a source is analysed to infer the properties of the lens. However, light emitted by the lens itself can contaminate the image of the source, introducing systematic errors in the analysis. We present a simple and efficient lens light model based on the well-tested multi-Gaussian expansion (MGE) method for representing galaxy surface brightness profiles, which we combine with a semi-linear inversion scheme for pixelized source modelling. Testing it against realistic mock lensing images, we show that our scheme can fit the lensed images to the noise level, with relative differences between the true input and best-fit lens light model remaining below 5\%. We apply the MGE lens light model to 38 lenses from the HST SLACS sample. We find that the new scheme provides a good fit for the majority of the sample with only 3 exceptions -- these show clear asymmetric residuals in the lens light. We examine the radial dependence of the ellipticity and position angles and confirm that it is common for a typical lens galaxy to exhibit twisting, non-elliptical isophotes and boxy~/~disky isophotes. Our MGE lens light model will be a valuable tool for understanding the hidden complexity of the lens mass distribution.   
\end{abstract}




\section{Introduction}
Galaxy-galaxy strong gravitational lensing is a phenomenon whereby light rays emitted from distant galaxies are strongly deflected by the gravitational potential of a foreground galaxy when the galaxies are aligned along the line of sight. Multiple distorted images of background galaxies are observed in this phenomenon, and particularly, an Einstein ring of distorted images may be observed when the galaxies are perfectly aligned. The first observation of galaxy-galaxy strong lensing dates back to the late 1970s, when a pair of strongly lensed quasars were observed \citep{Walsh1979}. Since then, over one thousand galaxy-galaxy strong lensing systems have been discovered \citep{Huang2021}. Furthermore, within this decade, newly launched space telescopes, such as the James Webb Space Telescope (JWST) and Euclid are expected to discover hundreds of thousands of new systems \citep{Collett2015}. These lenses are invaluable to explore fundamental cosmological questions, such as searches for the small dark matter substructures that might help distinguish different dark matter candidates, tests of gravity, and independent measurement of the Hubble constant \citep{Collett2018, Birrer2022, Vegetti2023}.

Early-type massive elliptical galaxies dominate the statistics of observed galaxy-galaxy strong lenses, for example, the Sloan Lens Advanced Camera for Surveys (SLACS) sample \citep{Bolton2006}. Studies of unlensed early-type galaxies reveal significant surface brightness variations, including twisted isophotes, radial ellipticity changes, and boxy~/~disky isophotes \citep{Hao2006, Kormendy2009}. \citet{Goullaud2018} perform photometric analysis of 35 nearby massive elliptical galaxies with comparable masses to the SLACS sample and see evidence for all forms of surface brightness variations. At high galaxy masses, independent kinematic analysis \citep{Krajnovic2011} suggests twisting and boxy isophotes may be a result of the underlying 3D mass distribution being triaxial. No study has yet performed a detailed photometric analysis of a sample of strong lenses in this way and this work therefore provides the first valuable insight.



Such complexity presented by a lens galaxy introduces significant challenges to lensing analyses, particularly the modelling of lens light. The standard approach for modelling lens light is to use multiple S\'ersic profiles. However, \citet{Nightingale2024} have demonstrated that S\'ersic profiles cannot accurately capture complexities observed in real lenses. Beyond the application of S\'ersic profiles, \citet{Bolton2006, Bolton2012} have adopted a B-spline interpolation model to fit and subtract the lens light in SLACS lenses. This model has shown flexibility in representing complex morphologies of lens light emission. However, because this model is not fitted simultaneously with the source model, it is necessary to manually mask out the lensed source region during the lens light fitting process. This requires prior knowledge of the location and intensities of the lensed source emission, a task that becomes particularly challenging with complex sources.

A multi-Gaussian expansion (MGE) method has been widely used in stellar dynamics analyses and has shown great success in fitting the surface brightness of realistic galaxies, especially ellipticals, with remarkable accuracy \citep{Cappellari2002, Cappellari2008, Li2019, Zhu2023}. The key idea is to approximate a galaxy's brightness using tens or hundreds of Gaussian profiles. As demonstrated by \citet{Cappellari2002}, this approach allows for significant flexibility because each Gaussian profile can be adjusted independently in terms of size, axis ratio, position angle and intensity, enabling the model to accurately reflect the varying ellipticity and twisted isophotes observed in real elliptical galaxies. However, the complexity and the huge number of adjustable parameters have limited the MGE model's application in strong lensing analyses. To effectively incorporate it into lensing modelling, a careful balance must be achieved between the model's adaptability and computational demands.

Beyond the complexity of a lens galaxy's lens light distribution, the lens mass complexity has recently become a focal point of discussion, driven by the increasing need for more precise lens mass models. The conventional lens mass model, a single elliptical power law profile \citep{Tessore2015}, has encountered notable limitations in various aspects of strong lensing analysis. This includes challenges in inferring the Hubble constant \citep{Cao2021, Van2021}, detecting subhaloes in strong lenses \citep{He2023, Nightingale2024} and measuring cosmic shear around lenses \citep{Etherington2023}. In an effort to surpass these limitations, researchers have introduced several sophisticated models, including both free-form mass models and those integrating additional multipole components \citep{Vegetti2009a, Riordan2023, Stacey2024, Cohen2024} or stellar disks \citep{Hsueh2017, Hsueh2018}. However, the full picture of lens mass complexity remains unclear. It is widely hypothesized that the systematic inaccuracies in current lens mass modelling may relate to the observed complexities in stellar light, such as twisted isophotes. This is because, in a standard galaxy-galaxy lensing scenario, the stellar mass, which closely traces the stellar light distribution, dominates the lens mass around and within the Einstein radius (several kiloparsecs). Therefore, refining the modelling of the lens light acts as a crucial step towards a comprehensive understanding and more precise modelling of lens mass complexity.

In this work, our objective is to build a lens light model that meets two key criteria. Firstly, the model should have the capability to represent the surface brightness of most real lens galaxies, accounting for twisting non-elliptical isophotes. Secondly, it should be simple enough so that it can be easily integrated into the existing framework for strong lensing analyses and therefore be fitted simultaneously with a model for the lens's mass and a pixelized source reconstruction. To achieve that, we propose a lens light model based on the MGE technique. The MGE model also offers the advantage of making an analytical computation of deflection angles possible, allowing for a straightforward conversion of the surface brightness representation into lensing quantities, assuming a specific mass-to-light ratio relation \citep{Shajib2019}. MGE mass modelling will be explored in our future work.

The MGE provides numerous benefits for automating strong lens analysis \citep{Etherington2022}, which is key with surveys such as Euclid poised to discover over 100,000 systems \citep{Collett2015}. This will be expanded upon in {Fran{\c{c}}a} et al. 2024, who perform automated analysis of 21 lenses using the MGE for not only the lens light but also the source galaxy.


The paper is structured as follows: In Section~\ref{sec:model}, we introduce our MGE lens light model, its implementation and the lensing modelling pipeline; In Section~\ref{sec:mock_test}, we test the MGE lens light model against mock datasets with realistic lens and source emission; In Section~\ref{sec:slacs_results}, we then apply the model to real HST observed lenses to test its performance on real data and show the complexity of the surface brightness of realistic lens galaxies; In Section~\ref{sec:discussion}, we discuss our results; Finally, in Section~\ref{sec:conclusions}, we summarize our results. Throughout the paper, unless otherwise specified, all lensing-related calculations are conducted using the public code \textsc{PyAutoLens}\footnote{\url{https://github.com/Jammy2211/PyAutoLens}} \citep{Nightingale2018, Nightingale2019, Nightingale2021, Nightingale2024}. For our analyses, we adopt a
spatially-flat $\Lambda$-CDM cosmology with H$_0=67.8 \pm 0.9\,\mathrm{km}\,\mathrm{s}^{-1}\,\mathrm{Mpc}^{-1}$ and $\Omega_{\rm M}=0.308 \pm 0.012$
\citep{Planck2016}.

\section{Model and Method}\label{sec:model}

\subsection{Overview}
\begin{figure*}
    \includegraphics[width=2.0\columnwidth]{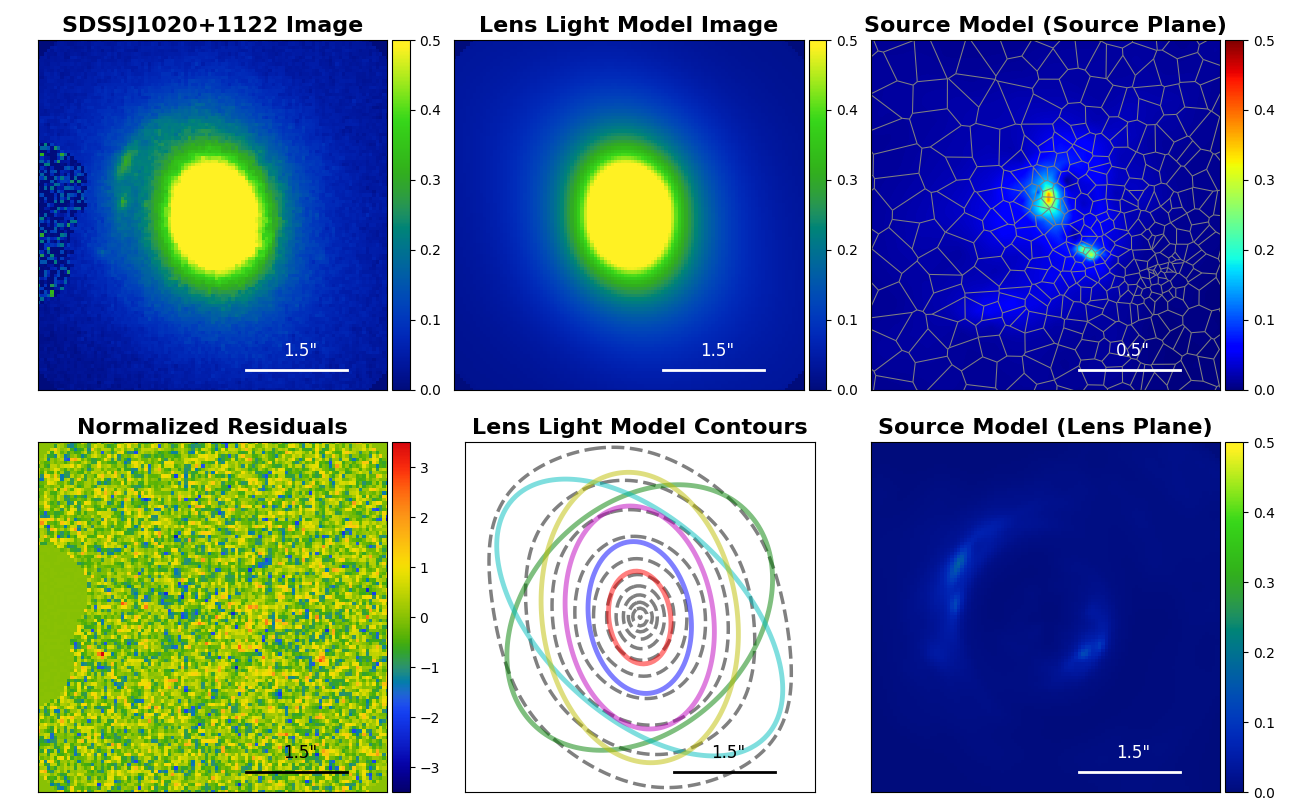}
    \caption{An overview of the lensing modelling of SDSSJ1020+1122 with the MGE lens light model described in this paper. The MGE lens light model used here consists of 6 sets of Gaussian profiles, where each set has 30 Gaussian profiles with sigmas spreading the image. \textbf{Top left:} The HST image of SDSSJ1020+1122 (a nearby galaxy on the left has been masked). \textbf{Top middle:} The best-fit MGE lens light model image. \textbf{Top right:} The reconstructed source image (on the source plane). The grey lines mark the Voronoi mesh used to construct the source plane. \textbf{Bottom left:} The normalized residuals of the best-fit model ((model - data) / errors). The colour bar ranges from -3.5 to +3.5. \textbf{Bottom middle:} Dashed grey lines are the isophotes of the best-fit MGE lens light model. Coloured solid lines mark (only) the position angles and axis ratios of the 6 sets of Gaussian profiles of the MGE lens light model. \textbf{Bottom right:} The reconstructed source image (on the lens plane).}
    \label{fig: overview}
\end{figure*}

An illustrative example of applying our MGE lens light model to a real HST observed image, SDSSJ1020+1122, is shown in Fig.~\ref{fig: overview}. The lens galaxy's emission is fitted using a total of 180 Gaussians, which are grouped into 6 sets of 30 Gaussians, where in each set the Gaussians share the same axis ratios, position angles and centre. The source galaxy is fitted simultaneously with the MGE, using a Voronoi mesh which reconstructs its irregular morphological appearance. The MGE fits all of the lens galaxy's emission and effectively captures its twisted isophotal contours, as illustrated in the bottom middle panel of Fig.~\ref{fig: overview}. More commonly adopted S\'ersic profiles cannot capture this complexity. The MGE has over $500$ parameters and the source Voronoi mesh over $1000$. This method section will outline how we parameterize both in a way that makes their joint application computationally tractable. Our MGE implementation can also be applied to non-lensed galaxies via the software \textsc{PyAutoGalaxy}\footnote{\url{https://github.com/Jammy2211/PyAutoGalaxy}}.

We now explain our models and methods in detail. We first describe the standard semi-linear method for fitting a pixelated model of the source light in \S\ref{sec:source_pixelization}. We then present the MGE model, designed for modelling the lens light, in \S\ref{sec:multiple_gaussian_expansion}. Next,  we detail the integration of the MGE lens light model into the likelihood optimisation framework of our strong lensing analysis in \S\ref{sec:implementing_mge}. For completeness, we briefly explain the exact source and lens mass models used in this work in \S\ref{sec:source_model} and \S\ref{sec:lens_mass_model}. Finally, in \S\ref{sec:modelling_pipelines}, we summarize our modelling pipelines used in \textsc{PyAutoLens}.

\subsection{Semi-linear scheme for modelling a pixelized source}\label{sec:source_pixelization}
To model background source galaxies of various morphologies, \citet{Warren2003} introduced a pixelized source model that represents the background source plane as a rectangular mesh with hundreds of grid points, each having its flux as a free parameter. With this source model, the lens fitting problem is posed as finding the best lens mass parameters and fluxes for the source plane pixels that can fit the lensed image with realistic source emission. Here, realistic implies the source emissions should resemble physical components with spatially smooth fluxes.

A semi-linear inversion (SLI) method is used to solve the fitting problem. For a lens model (without any lens light), the parameters are naturally divided into two categories: the fluxes of source plane pixels and the lens mass parameters. The lens mass parameters are non-linear, meaning that they can only be inferred by non-linear fitting, for example by a nested sampling algorithm. The source parameters (or equivalently pixels) are linear, whereby their flux values can be efficiently solved for via a linear inversion. The combination of non-linear and linear parameters gives it the name ``semi-linear''.

Given a set of lens mass parameters, one can build up a linear mapping relation between the fluxes of the source plane grid, $S_{j}\, (j=1, 2, ..., N_S)$, and the fluxes of the model image grid, $I_{i}\, (i=1, 2, ..., N_I)$, as
\begin{equation}\label{equ: mapping_function}
    I = f \times S,
\end{equation}
where $f$ is the (source-lens plane) mapping matrix and an element, $f_{ij}$, quantifies the fraction of the flux contributed from the $j$-th source grid to the $i$-th image grid. $f_{ij}$ is determined by the ray-tracing of the lens mass profile, the source pixelization and instrumental effects, such as blurring due to the point spread function (PSF).

With the mapping relation, Eq.~\eqref{equ: mapping_function}, the $\chi^2$ is then defined as
\begin{subequations}\label{equ: lensing_img_chi2}
    \begin{align}
        \chi^2 & \equiv \frac{1}{2}\sum_{i=1}^{N_I}\left[\frac{I_i - d_i}{n_i}\right]^2 \label{equ: lensing_img_chi2_l1} \\
               & = \frac{1}{2} \sum_{i=1}^{N_I}\left[\frac{\sum_{j=1}^{N_S} f_{ij}S_j - d_i}{n_i}\right]^2 \label{equ: lensing_img_chi2_l2} \\
               & = \frac{1}{2}||ZS - Y||^2, \label{equ: lensing_img_chi2_l3}
    \end{align}
\end{subequations}
where
\begin{subequations}
    \begin{align}
        Z_{ij} & \equiv F_{ij} / n_i, \label{equ: lensing_img_chi2_l4} \\
        Y_i & \equiv d_i / n_i.\label{equ: lensing_img_chi2_l5}
    \end{align}
\end{subequations}
The image data is $d_i$ and the noise is $n_i$. In terms of $S$, Eq.~\eqref{equ: lensing_img_chi2_l3} has a quadratic form, confirming that the $\chi^2$ minimization problem reduces to a classical least-squares fitting problem with respect to source fluxes. If we have no boundary constraints on the solutions, the best-fit source fluxes, $\hat{S}$, can be analytically expressed as
\begin{equation}
    \hat{S} = \left(Z^{\rm T}Z\right)^{-1}Z^{\rm T}Y.
\end{equation}
The time cost for solving $\hat{S}$ is small, which makes it feasible to reconstruct a source galaxy with thousands of pixels.

We also require our model to produce a realistic background source. A regularization term, $G_L$, is therefore added to the goodness function, $G$, as
\begin{equation}\label{eq: lensing_img_chi2}
    G = \chi^2 + G_L .
\end{equation}
To ensure that $G$ keeps the quadratic form as Eq.~\eqref{equ: lensing_img_chi2_l3}, we require $G_L$ to be a quadratic polynomial function of $S_i (i=1, 2, ..., N_S)$ as
\begin{equation}
    G_L \equiv \frac{1}{2}r\sum_{i, j} H_{ij} S_i S_j,
\end{equation}
where $r$ is the regularization strength and $H_{ij}$ is called the regularization matrix, whose elements are determined by the choice of the regularization scheme. By design, the regularization matrix is a positive-definite matrix, and thus has a ``square root'' matrix, $B_{ij}$, such that $H = B^TB$. Combining Eq.~\eqref{eq: lensing_img_chi2} and the regularization term, the goodness function, Eq.~(\ref{eq: lensing_img_chi2}), can be expressed as 
\begin{subequations}\label{equ: lensing_goodness}
    \begin{align}
        G & = \chi^2 + G_L \label{equ: lensing_goodness_l1} \\
          & = \frac{1}{2}\left\Vert\left(\begin{array}{c}Z \\ \sqrt{r}B \end{array}\right)\cdot S  - \left(\begin{array}{c}Y \\ 0\end{array}\right)\right\Vert^2, \label{equ: lensing_goodness_l2}
    \end{align}
\end{subequations}
which keeps the quadratic form. 

Following the descriptions of \citet{Suyu2006, Nightingale2015}, to objectively determine the regularization strength, we put the SLI procedure in a Bayesian framework which maximizes the Bayesian evidence, $\epsilon$, defined as
\begin{equation}\label{equ: bayesian_evidence}
    \begin{split}
   - 2\ {\rm ln}\ \epsilon \equiv \chi^2 + & G_L + {\rm ln}\left[{\rm det}\left(f^Tf + r H\right)\right] -
   {\rm ln}\left[{\rm det}\left(r H\right)\right] \\ & + \sum_{i=1}^{N_I}{\rm ln}\left[2{\rm \pi}n_i^2\right].
   \end{split}
\end{equation}


\subsection{Multiple-Gaussian Expansion}\label{sec:multiple_gaussian_expansion}

\subsubsection{Basic formulas}
The MGE model represents a galaxy's surface brightness, $I(x, y)$, by a set of independent 2D Gaussian profiles of the form
\begin{equation}
    I(x, y) = \sum_{i}^{N} G_i(x, y).
\end{equation}
Here, $G_i$ is the $i$-th Gaussian profile 
\begin{equation}
    G_i(x, y) = I_{i}\cdot{\rm \exp}\left(-\frac{R_i^2(x, y)}{2\sigma_i^2}\right),
\end{equation}
where $I_i$ and $\sigma_i$ are the $i$-th Gaussian profile's intensity and width. $R_i(x, y)$ is its elliptical radius defined as
\begin{equation}
    \begin{split}
        & R_i(x, y) = \sqrt{x^{\prime2} + \left(\frac{y^\prime}{q_i}\right)^2} , \\
        & x^{\prime} = \cos{\theta_i}\cdot\left(x - x^{\rm c}_i\right) + \sin{\theta_i}\cdot\left(y - y^{\rm c}_i\right) , \\
        & y^{\prime} = \cos{\theta_i}\cdot\left(y - y^{\rm c}_i\right) -\sin{\theta_i}\cdot\left(x - x^{\rm c}_i\right),
    \end{split}\label{eq: ellp}
\end{equation}
where $q_i$, $\theta_i$, $\left(x^{\rm c}_i,\ y^{\rm c}_i\right)$ are the profile's axis ratio, position angle and centre, respectively. A 2D Gaussian profile has 6 parameters, and an MGE model of $N$ Gaussians has $6 \times N$ parameters.

\subsubsection{MGE parameterization}
To simplify the MGE model for lens modelling and reduce the number of non-linear free parameters, we introduce the concept of a ``set''. A set is a group of Gaussians which share the same centre, position angle, and axis ratio, and with $\sigma$ values fixed to preset values. We typically fix the $\sigma$ values evenly in log-spaced intervals between one-fifth of the image pixel scale and the circular radius of the mask applied to the lens image. 
Our overall MGE lens light model is then a combination of multiple sets of Gaussians with different centres, position angles, and axis ratios. The formula is then written as
\begin{equation}
    \begin{split}
    I^{\rm lens}_{\rm MGE}(x, y) & = \sum^{N_{\rm b}}_{i=1}B_{i}(x, y) \\
    & = \sum^{N_{\rm b}}_{i=1}\sum^{N_{\rm g}^{i}}_{j=1}G_{i, j}(x, y),
    \end{split}
\end{equation}
where $B_{i}$ is the $i$-th set of Gaussians, $N_{\rm b}$ is the number of sets and $N_{\rm g}^i$ is the number of individual Gaussians of the $i$-th set. One can simply customize $N_{\rm b}$ and $N_{\rm g}^i$ to change the flexibility of the model depending on the complexity of the surface brightness of a lens galaxy. 

To model the extended luminous emission in each lens galaxy, we assume $N_{\rm g}^i = 30$ and $N_{\rm b} = 2$, 4 or 6, where these values depend on the complexity of each lens. We also perform fits including a ``point'' set, which fits point-source emission found in the centre of a subset of lenses. A point set consists of 10 Gaussians, has $\sigma$ values spanning one-fifth the pixel size to two times the pixel size and has a centre that is independent of the centres of the other sets that model the extended emission.   

\subsubsection{Semi-linear solution for Gaussian intensities}

We now describe the semi-linear procedure which solves for the intensity of each Gaussian linearly, further reducing the number of non-linear parameters of our MGE model. The surface brightness of the MGE model is a linear combination of the intensity of each Gaussian, therefore the intensities can be linearly solved once their other parameters (centres, axis ratios, sigma values) are given. Let us first consider fitting the MGE model to an image without any lensing. Given an image, $d$, with $N_{\rm pix}$ pixels and a noise map, $n$, we minimize the $\chi^2$ of the fit defined as
\begin{subequations}\label{equ: mge_chi2}
\begin{align}
    \chi^2 & \equiv \frac{1}{2}\sum_i^{N_{\rm pix}}\left[\frac{I_{\rm MGE}^{\rm lens}(x_i, y_i) - d_i}{n_i}\right]^2 \label{equ: mge_chi2_l1} \\
           & = \frac{1}{2}\sum_i^{N_{\rm pix}}\left[\frac{\sum_{j}^{N_{\rm g}^{\rm tot}}I_j\cdot A_{ij} - d_i}{n_i}\right]^2 \label{equ: mge_chi2_l2} \\
           & = \frac{1}{2}||X I - Y||^2, \label{equ: mge_chi2_l3}
\end{align}
\end{subequations}
where
\begin{subequations}
    \begin{align}
    & A_{ij} \equiv {\exp}\left(\frac{-R^2_{i}(x_i, y_i)}{2\sigma_{i}^2}\right) \label{equ: mge_chi2_l4}\\
    & X_{ij} \equiv A_{ij} / n_i \label{equ: mge_chi2_l5} \\
    & Y_{i} \equiv d_i / n_i. \label{equ: mge_chi2_l6} 
    \end{align}
\end{subequations}
$\chi^2$ can again be expressed as a quadratic form with respect to the intensities of Gaussians. As a result, we can obtain the best-fit intensities of the Gaussians linearly. 

This formalism ensures that the intensities of the Gaussians are not fitted for by the non-linear sampler. Therefore, only the centres, axis ratios and position angles of each set are. For example, for an MGE model composed of 3 sets, where each set has 30 Gaussians, it has just $3\times4=12$ non-linear free parameters. By assuming the Gaussians in all sets share the same centre (which is a common assumption for realistic galaxies), this reduces to $2 + 2\times3=8$ parameters. This is less than the number of parameters used to fit S\'ersic models \citep{Nightingale2024}, which have significantly less flexibility than the MGE.

\subsection{Solving for the MGE and pixelized source simultanoeusly}\label{sec:implementing_mge}

\subsubsection{Semi-linear solution for both lens and source intensities}

We now describe how using the semi-linear inversion method we solve for the MGE lens light intensities and pixelized source pixel fluxes simultaneously. We extend the goodness function, Eq.~(\ref{equ: lensing_goodness}), as
\begin{subequations}\label{equ: lensing_mge_goodness}
    \begin{align}
        G & = \chi^2 + G_{\rm L} + G_{\rm M} \label{equ: lensing_mge_goodness_l1} \\
          & = \frac{1}{2}\left(\sum_{i=1}^{N_{\rm I}}\left[\frac{\sum_{j=1}^{N_{\rm s}}f_{ij}S_{j} + \sum_{k=1}^{N_{\rm g}}I_kA_{ik} - d_i}{n_i}\right]^2 + G_{\rm L} + G_{\rm M}\right) \label{equ: lensing_mge_goodness_l2} \\
          & = \frac{1}{2}\left|\left|\left(
          \begin{array}{cc}
               Z & X \\
               \sqrt{r}B & B_{\rm M} 
          \end{array}
          \right)\cdot \left(
          \begin{array}{c}
               S  \\
               I
          \end{array}
          \right) - 
          \left(\begin{array}{c}
               Y  \\
               0 
          \end{array}\right)
          \right|\right|^2, \label{equ: lensing_mge_goodness_l3}
    \end{align}
\end{subequations}
where $G_M$ is a regularization term for the MGE light model, and $B_{\rm M}$ is the corresponding ``square root'' of the regularization matrix. The MGE regularization stabilizes the quadratic minimization process but has a negligible impact on the inferred intensity values. We set it to be the identity matrix.

\subsubsection{Non-negative least-square solver}
Without any boundary constraints on the solutions, $\left(\begin{array}{c}S \\ I \end{array}\right)$, the results can be analytically obtained by setting the first derivative of Eq.~\eqref{equ: lensing_mge_goodness_l3} to be zero, which is a common choice for pixelized source models \citep{Warren2003, Nightingale2018}. However, in Section~\ref{sec: mock_results} we show that by combining two ``free-form'' models (both MGE lens light and pixelized source), a significant degeneracy arises between the two, particularly around the lensed arc regions where both models are capable of fitting the light. This can bias the recovery of the lens and source emission and produce overfitting whereby the intensities of the Gaussians alternate between large positive and negative values.  

We therefore require our solution (both the intensities of Gaussian profiles and the fluxes of source pixels) to be non-negative, which is a well-posed physical assumption as both the source and lens galaxies should not have negative fluxes in reality. The trade-off for switching to a non-negative SLI scheme is slower computational run time. To alleviate this performance decrease, we use a modified version of the fast non-negative least-square (fnnls) algorithm to minimize Eq.~(\ref{equ: lensing_mge_goodness_l3}) \citep{Bro1997}\footnote{The fnnls code we are using is modified from \url{https://github.com/jvendrow/fnnls}.}. Hereafter, we use ``nn-MGE'' to refer to analysis using an MGE lens light model with non-negative constraints (on both the MGE lens light and pixelized source model) and ``pn-MGE'' to refer to analysis without the non-negative constraints (on both the MGE and pixelized source). 

\subsubsection{Bayesian framework}
This new goodness of fit is incorporated into an extension of the Bayesian framework shown in Eq.~(\ref{equ: bayesian_evidence}), where
\begin{equation}\label{equ: bayesian_evidence_MGE}
    \begin{split}
   - 2\ {\rm ln}\ \epsilon \equiv \chi^2 + & G_L + G_{\rm M} + {\rm ln}\left[{\rm det}\left(f^Tf + r H\right)\right] -
   {\rm ln}\left[{{\rm det}\left(r H\right)}\right] \\ & + \sum_{i=1}^{N_I}{\rm ln}\left[2{\rm \pi}n_i^2\right]
   \end{split}.
\end{equation}
Here, $f$ and $H$ are the same as defined previously for a pixelized source model. 

\subsection{Source Implementation}\label{sec:source_model}

The formalism above describes how source pixel flux values are solved for via a mapping matrix $f$ and regularization matrix $H$. The construction of these matrices depends on the source implementation. In this work: (i) the source pixelization uses a Voronoi mesh with natural neighbour interpolation; (ii) the Voronoi cell centres are computed in the image plane via a k-means clustering algorithm and ray traced to the source plane via the mass model and; (iii) the regularization scheme adapts the degree of smoothing to the reconstructed source's luminous emission and interpolates values at a cross of surrounding points. The implementation details are given fully in Appendix~\ref{appxA}. This Appendix also illustrates how these features address the noisy and stochastic likelihood systematics described by \citet{Etherington2022}.

\subsection{Light and Mass}\label{sec:lens_mass_model}

A subset of fits will assume an elliptical S\'ersic profile, which is given by
\begin{equation}\label{equ:sersic}
I(r) = I^{'}{\rm exp}\left[- b_n \left(\frac{r}{r_{\rm e}}\right)^{1/n}\right],
\end{equation}
where $I^{'}$ is the intensity, $r_{\rm e}$ is the effective radius, $n$ is the S\'ersic index and $b_n$ is a normalizing coefficient determined by $n$ \citep{Graham2005}. Ellipticity is introduced to the profile following Eq.~(\ref{eq: ellp}).

The lens galaxy mass assumes the elliptical power-law profile \citep{Tessore2015}, whose convergence is described as
\begin{equation}
    \kappa(R) =
    \frac{3-\gamma}{1+q}\left(\frac{R_{\rm E}}{R}\right)^{\gamma-1},
 \label{equ:kappa_pl}
\end{equation}
where $R_{\rm E}$ is the Einstein radius, $q$ is the axis ratio and $\gamma$ is the slope. The ellipticity to the profile is introduced following Eq.~(\ref{eq: ellp}). The Singular Isothermal Ellipsoid (SIE) is the case where the power-law slope, $\gamma$, equals 2. 

An external shear is also included in the mass model, which is parameterized as two components, $\gamma^{\rm ext}_1$ and $\gamma^{\rm ext}_2$,
\begin{equation}\label{eq:shear}
    \gamma^{\rm ext} = \sqrt{\gamma_{\rm 1}^{\rm ext^{2}}+\gamma_{\rm 2}^{\rm ext^{2}}}, \, \,
     \tan{2\phi^{\rm ext}} = \frac{\gamma_{\rm 2}^{\rm ext}}{\gamma_{\rm 1}^{\rm ext}},
\end{equation}
where $\gamma^{\rm ext}$ is the shear strength and $\phi^{\rm ext}$ is the position angle of the shear.

\subsection{Lens modelling pipelines}\label{sec:modelling_pipelines}

We construct automated strong lens modelling pipelines using the SLaM (source, lens and mass) scripts provided by \textsc{PyAutoLens} \citep{Cao2021, Etherington2022, He2023, Nightingale2024}. These pipelines iteratively fit combinations of light, mass, and source models, where the initial stages fit simpler lens models (e.g., SIE mass profile, MGE source) for efficient and robust convergence towards accurate results and later stages employ more complex models (e.g., power-law mass model, Voronoi source reconstruction). We use non-linear nested samplers, \textsc{Dynesty} \citep{Speagle2020, Sergey2023} and \textsc{Nautilus} \citep{Lange2023}. 

\begin{table}
\renewcommand{\arraystretch}{1.5}
\centering
\begin{tabular}{ccccc}
\hline
\textbf{Pipeline} & \textbf{Phase} & \textbf{Component} & \textbf{Model} & \textbf{Prior info}
\\ \hline
\multirow{3}{*}{\begin{tabular}[c]{@{}c@{}}Source\\ Parametric\end{tabular}} & \multirow{3}{*}{\textbf{SP}}  & Lens mass & SIE + Shear & - \\ 
\cline{3-5} &  & Lens light & MGE & - \\ 
\cline{3-5} &  & Source light & MGE & - \\
\hline
\multirow{6}{*}{\begin{tabular}[c]{@{}c@{}}Source\\
Pixelized\end{tabular}}
& \multirow{3}{*}{\textbf{SPix1}} & Lens mass & SIE + Shear & \textbf{SP} \\
\cline{3-5} & & Lens light & MGE & \textbf{SP} \\
\cline{3-5} & & Source light & Voronoi & - \\
\cline{2-5} & \multirow{3}{*}{\textbf{SPix2}} & Lens mass & SIE + Shear & \textbf{SPix1} \\ \cline{3-5} & & Lens light & MGE & \textbf{SP} \\
\cline{3-5} & & Source light & Voronoi & - \\
\hline
\multirow{3}{*}{\begin{tabular}[c]{@{}c@{}}Light \end{tabular}}  & \multirow{3}{*}{\textbf{L}}  & Lens mass & SIE + Shear & \textbf{SPix1} \\
\cline{3-5} & & Lens light & MGE & \textbf{SP} \\
\cline{3-5} & & Source light & Voronoi & \textbf{SPix2} \\
\hline
\multirow{3}{*}{\begin{tabular}[c]{@{}c@{}}Mass\end{tabular}} & \multirow{3}{*}{\textbf{M}}  & Lens mass & EPL + Shear & \textbf{SPix1} \\
\cline{3-5} & & Lens light & MGE & \textbf{L} \\
\cline{3-5} & & Source light & Voronoi & \textbf{SPix2} \\
\hline
\end{tabular}
\caption{The Source, Light and Mass (SLaM) pipelines used in this analysis, built using \texttt{PyAutoLens}.}
\label{tab:pipeline_table}
\end{table}

Table \ref{tab:pipeline_table} provides an overview of each fit performed in the pipeline. The SLaM pipelines used in this work consist of five steps, a reduction from the ten or more used in previous works \citep{Etherington2022, Cao2021, Nightingale2024}. This reduction is made possible because the MGE simplifies the fitting of lens models in the early stages of the SLaM pipeline, which will be expanded upon in {Fran{\c{c}}a} et al. 2024.

An overview of the SLaM pipelines is as follows:
\begin{itemize}
    \item \textbf{Source Parametric Pipeline (SP)}:
Computes an accurate initial estimate of the lens model parameters. The lens and source light are modeled using an MGE and the lens mass uses an SIE with shear. The MGE lens light model comprises 2 sets of 30 Gaussian profiles, where all Gaussians share the same centre, but the two sets have different position angles and axis ratios. The MGE source uses one set of 20 Gaussians. Fits investigating S\'ersic profiles for the lens light use two S\'ersic profiles with the same centre. 
\item \textbf{Source Pixelized Pipeline (SPix)}:
     The source model is then made more complex, by using the Voronoi mesh. This is performed over two fits. For the first fit, the Voronoi mesh centres are based on the mass model magnification, whereas for the second fit, they adapt to the source's unlensed morphology. Both fits subtract the lens light using a 2-set MGE, where the centres, axis ratios and position angles of all Gaussians are fixed to the previous maximum likelihood solution. However, their intensities are still solved via the fnnls. For fits using S\'ersic lens light model, the light model is fully fixed to the previous maximum likelihood solution. The mass model of the first fit is an SIE plus shear where all parameters are free (with priors based on the inferred model from {\bf SP}). For the second fit, the mass model is fixed and only the source pixelization parameters are fitted for (the MGE intensities are again still solved via the fnnls). 

    \item \textbf{Light Pipeline (L)}:
    The Voronoi source reconstruction enables a cleaner deblending of the lens and source light. However, the lens light's non-linear parameters were previously fixed. The light pipeline therefore refits the MGE lens light model, assuming broad uniform priors on the parameters of the MGE sets. The mass model and source parameters are fixed to those inferred at the end of the \textbf{SPix} pipeline, however, the source fluxes are always solved simultaneously with the MGE via the fnnls. Different lens light models are fitted, including fits using 2, 4 or 6 sets of Gaussians, fits with and without a point-source emission set of 10 Gaussians and a model with three S\'ersic profiles. 
    
    \item \textbf{Mass Pipeline (M)}:
    The mass pipeline fits a more complex lens mass distribution, the single elliptical power law (EPL) plus shear. The parameters of the lens light MGE and source reconstruction are fixed to those inferred in the light and source pipelines. However, their intensities are linearly solved for, for every mass model fitted. When using the S\'ersic light model, we simultaneously fit the light model parameters (e.g. effective radii, S\'ersic index) alongside the mass model. The \textbf{Mass Pipeline} is the final step of our lensing modelling procedure in this work. All the results reported in this work are the results of this pipeline.
\end{itemize}

\begin{figure*}
    \includegraphics[width=2.0\columnwidth]{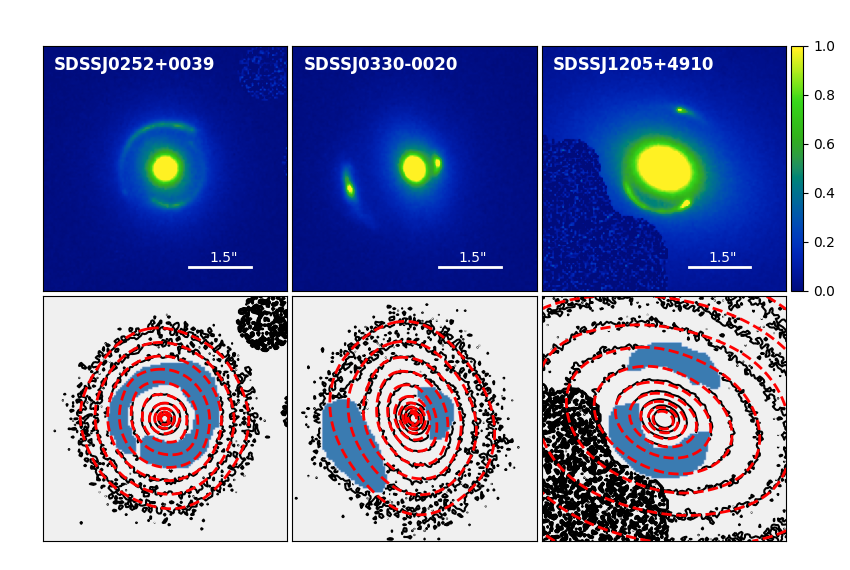}
    \caption{\textbf{Top row:} The HST observations of three SLACS lenses, SDSSJ0252+0039, SDSSJ0330-0020 and SDSSJ1205+4910. Contaminations of nearby galaxies are masked out. \textbf{Bottom row:} The isophotes of real observations (black) and the best-fit model of \textsc{mgefit} (red). The blue patterns mark the regions of source emissions, where are masked out when fitted by \textsc{mgefit}.}
    \label{fig:mge_mocks}
\end{figure*}

\section{Tests on Mock data}\label{sec:mock_test}
In this section, we first simulate three mock lensing images with realistic lens and source emission. We then use these simulated images to test the MGE lens light model and compare its results with those obtained using a S\'ersic lens light model. 

\begin{figure*}
    \includegraphics[width=2.0\columnwidth]{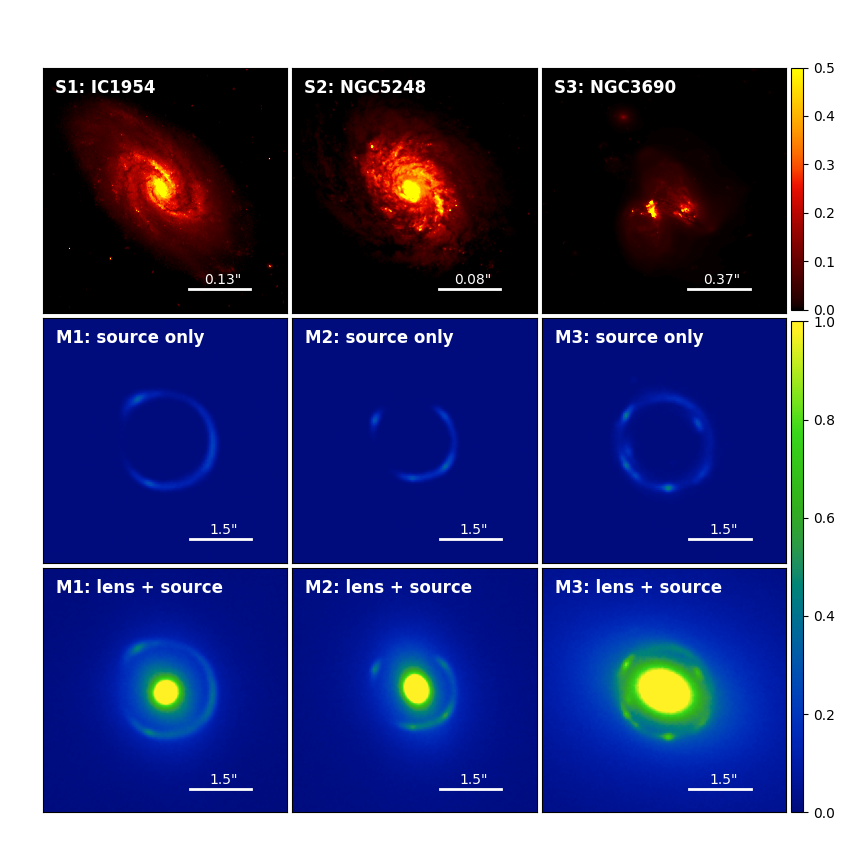}
    \caption{\textbf{Top row:} HST images of three nearby galaxies that are used as source galaxies in our mocks. The angular scale is indicated in the bottom right corner of each panel. \textbf{Middle row:} Source-only images of three mocks. \textbf{Bottom row:} Three mock images were used for our tests.}
    \label{fig:mock_imgs}
\end{figure*}

\begin{table*} 
    \centering
    \begin{tabular}{ccccccccccccccccc} 
    \hline
    \multicolumn{17}{c}{\textbf{SDSSJ0252+0039}} \\
    \hline
     & g1 & g2 & g3 & g4 & g5 & g6 & g7 & g8 & g9 & g10 & g11 & g12 & g13 & g14 & g15 & g16 \\
    $q$ & 0.83 & 0.77 & 0.83 & 0.90 & 0.65 & 0.64 & 0.77 & 0.50 & 1.00 & 0.79 & 0.48 & 0.62 & 0.41 & 0.18 & 0.78 & 0.22 \\
    $\phi$ & -19 & -19 & 71 & -19 & -70 & 4 & 56 & -52 & 0 & -80 & -2 & -78 & 28 & 49 & -73 & -25 \\
    lg~$I$ & 0.86 & 0.91 & 0.29 & 0.30 & -0.24 & -0.48 & -0.13 & -0.75 & -0.73 & -0.76 & -1.6 & -1.4 & -2.0 & -2.2 & -2.2 & -2.4 \\
    lg~$\sigma$ & -1.8 & -1.6 & -1.4 & -1.1 & -1.0 & -0.85 & -0.75 & -0.56 & -0.25 & -0.14 & -0.13 & 0.0018 & 0.020 & -0.19 & 0.43 & -0.11 \\
    \hline
    \multicolumn{17}{c}{\textbf{SDSSJ0330-0020}} \\
    \hline
    & g1 & g2 & g3 & g4 & g5 & g6 & g7 & g8 & g9 & g10 & g11 & g12 & g13 & g14 & g15 & g16 \\
    $q$ & 0.82 & 0.64 & 1.0 & 0.60 & 0.62 & 0.91 & 0.58 & 0.34 & 1.00 & 0.46 & 0.59 & 0.76 & 0.21 & 0.30 & 0.44 & 0.48 \\
    $\phi$ & 65 & -64 & 0 & -78 & -23 & 65 & -55 & 84 & 0 & -66 & -89 & -40 & -81 & 7 & 65 & -47 \\
    lg~$I$ & 1.0 & 0.69 & 0.26 & 0.14 & -0.47 & -0.19 & -0.63 & -1.1 & -1.1 & -0.86 & -1.1 & -1.1 & -2.3 & -2.3 & -1.8 & -1.7 \\
    lg~$\sigma$ & -1.7 & -1.4 & -1.1 & -1.1 & -0.91 & -0.72 & -0.72 & -0.78 & -0.25 & -0.50 & -0.23 & -0.11 & -0.15 & 0.018 & 0.18 & 0.22 \\
    \hline
    \multicolumn{17}{c}{\textbf{SDSSJ1205+4910}} \\
    \hline
    & g1 & g2 & g3 & g4 & g5 & g6 & g7 & g8 & g9 & g10 & g11 & g12 & & & & \\
    $q$ & 0.92 & 1.0 & 0.85 & 0.62 & 0.84 & 0.70 & 0.73 & 0.66 & 0.58 & 0.67 & 0.36 & 0.84 & & & & \\
    $\phi$ & -63 & -90 & -24 & 27 & -62 & -25 & -18 & -70 & -25 & 8 & -31 & -28 & & & & \\
    lg~$I$ & 0.78 & 1.1 & 0.71 & -0.65 & -0.24 & 0.077 & -0.015 & -1.1 & -0.61 & -0.90 & -1.4 & -1.2 & & & & \\
    lg~$\sigma$ & -1.8 & -1.3 & -0.98 & -0.86 & -0.68 & -0.75 & -0.44 & -0.22 & -0.17 & 0.085 & 0.082 & -1.2 & & & & \\
    \hline
    \end{tabular}
    \caption{Best-fit \textsc{mgefit} results of the three selected SLACS lenses. The units of $x$, $y$, $\sigma$ is arcsec. The units of $I$ is $e^-{\rm s}^{-1}{\rm pix}^{-1}$ and the units of $\phi$ is degrees. The centre (x, y) of the three best-fit MGE light models are (0.013\arcsec, -0.070\arcsec), (0.048\arcsec, 0.017\arcsec) and (-0.001\arcsec, -0.034\arcsec) respectively for SDSSJ0252+0039, SDSSJ0330-0020, and SDSSJ1205+4910.}\label{tab:mgefit_results}
\end{table*}

\subsection{Mock Data}
To create mock lensing data with lens light emission close to that of real lens galaxies, we simulate the lens light of three lensing systems from the SLACS sample \citep{Bolton2006}, SDSSJ0252+0039, SDSSJ0330-0020, and SDSSJ1205+4910. We need to simulate images containing the complex lens emission our MGE is designed to fit, but also want to simulate images independently of our specific MGE implementation. We therefore use \textsc{mgefit}  \citep[a public MGE fitting code by][]{Cappellari2002} to fit these real HST strong lensing images, where the source emission is manually masked out. \textsc{mgefit} models the emission of a galaxy using multiple Gaussian profiles. However, unlike our MGE model, \textsc{mgefit} does not group the Gaussian profiles into sets and each Gaussian profile has its own position angle and axis ratio. When using \textsc{mgefit}, we assume all the Gaussian profiles share the same centre. 

In the first row of Fig.~\ref{fig:mge_mocks}, we show the HST images of the three selected SLACS systems, and in the bottom row, we compare the best-fit \textsc{mgefit} results (red dashed contours) with the observations (black contours). \textsc{mgefit} fits the lens emission accurately, capturing the complex features we want to include in our lensing mocks, such as the twisting of isophotes. The blue patterns mark the source emission regions, which are not included in the \textsc{mgefit} fitting procedure. We use the best-fit \textsc{mgefit} results to represent the lens light emission of our mock datasets. The best-fit \textsc{mgefit} results are listed in Table~\ref{tab:mgefit_results}. In total, the lens light of SDSSJ0252+0039, SDSSJ0330-0020, and SDSSJ1205+4910 are represented, respectively, by 16, 16, and 12 individual Gaussian profiles.

\begin{table*}
    \centering
    \begin{tabular}{cccccccccc} 
    \hline
    & lens light & source light & mass ($x$, $y$)~[($\arcsec$, $\arcsec$)] & mass $R_{\rm E}~$[$\arcsec$] & mass $q$ & mass $\phi$~[$^\circ$] & mass $\gamma$ & $z_{\rm l}$ & $z_{\rm s}$ \\
    \hline
    M1 & SDSSJ0252+0039 & S1: IC1954 & (0.013, -0.070) & 1.2 & 0.8 & 45 & 2.1 & 0.28 & 0.98 \\
    M2 & SDSSJ0330-0020 & S2: NGC5248 & (0.048, 0.017) & 1.0 & 0.77 & 60 & 1.9 & 0.35 & 1.1 \\
    M3 & SDSSJ1205+4910 & S3: NGC3690 & (-0.001, -0.034) & 1.1 & 0.75 & 30 & 2.0 & 0.22 & 0.48 \\
    \hline
    \end{tabular}\caption{Parameters used for the three mocks.}\label{tab:mock_paras}
\end{table*}

For the mock source galaxies, we use HST images of three nearby galaxies, IC1954, NGC5248, and NGC3690 to ensure that our mock data have source galaxies with complex structures. The images of these three galaxies are shown in the top row of Fig.~\ref{fig:mock_imgs}. We obtain an unrealistic extended lensed arc (compared to real lens observations) if we directly place the images of the nearby galaxies at the source redshifts. This is because actual strong lens source galaxies, such as SLACS, are more compact than the three nearby galaxies we selected here. Consequently, to simulate lensed arcs with similar sizes as those seen in real observations, we re-scale (shrink) the size of the images when placing them at the source redshifts. Their actual sizes, used in our mocks, are denoted in the bottom right corner of each panel. Two of them are spiral galaxies with clear arm features, while the third is a system of two merging galaxies.  

As the main goal here is testing our MGE lens light model, we simply set the mock lens mass as an elliptical power-law profile as Eq.~(\ref{equ:kappa_pl}). We add no external shear to the lens mass. However, when we model the mock images we include external shear as part of our model.

We simulate our mock images as HST observations with a pixel size of $0.05\arcsec$ and a Gaussian PSF with a sigma size equivalent to the pixel size. For a given mock lens mass, following the lensing equation, we trace each image pixel back to the source plane to retrieve the flux value of the lensed source image. To ensure that our lensed images capture sufficient details of the source galaxies, we sample each image pixel with $16\times16$ sub-pixels. This means that for each image pixel, we trace back 256 light rays to the source galaxy plane and take the average flux of these 256 light rays as the value of that image pixel. We then add the lens light emission to the lensed images and convolve the resulting images with the PSF to create mock lensing images. We set our mock images to match the data quality, resembling the best case of SLACS lenses, which has a maximum pixel signal-to-noise (S/N) ratio $\sim50$. 

In total, we simulate three mock lensing images denoted as M1, M2, and M3: the lens light of M1 is from SDSSJ0252+0039 and the source is IC1954; the lens light of M2 is from SDSSJ0330-0020 and the source is NGC5248; the lens light of M3 is from SDSSJ1205+4910 and the source is NGC3690. In the bottom row of Fig.~\ref{fig:mock_imgs}, we show the simulated mock lensing images. For clarity, we also show the source-only versions in the middle row. In Table~\ref{tab:mock_paras}, we summarize the parameters used for the three mocks.

\subsection{Mock test results}\label{sec: mock_results}
We now use the SLaM pipeline to fit the mock imaging. The lens mass model uses the elliptical power law profile plus an external shear (the same model used to simulate the mocks) and the source uses the Voronoi mesh. In the \textbf{Light Pipeline}, when using the MGE light model, we use 4 sets of Gaussians, each composed of 30 Gaussian profiles, all with the same centre. The additional set of 10 Gaussian profiles which model central point source emission in the lens is also included (see Section~\ref{sec:point_centre}). We first perform fits enforcing non-negative constraints on the solutions of source pixel fluxes and Gaussian intensities (denoted as ``nn-MGE''), which we compare to fits without boundary conditions for our solutions (as ``pn-MGE''). We also compare to models where the lens light is fitted using three S\'ersic profiles, which share the same centres but have different position angles, axis ratios, and other parameters. We do not constrain the source pixel fluxes to be positive-only when using the S\'ersic model. We fit data within circular masks containing the whole lensed arcs. For mock M1 and M3, the mask radius is $1.8\arcsec$ and the mask radius of M2 is $1.6\arcsec$ because it has a slightly smaller Einstein radius.

\begin{figure*}
    \includegraphics[width=2.0\columnwidth]{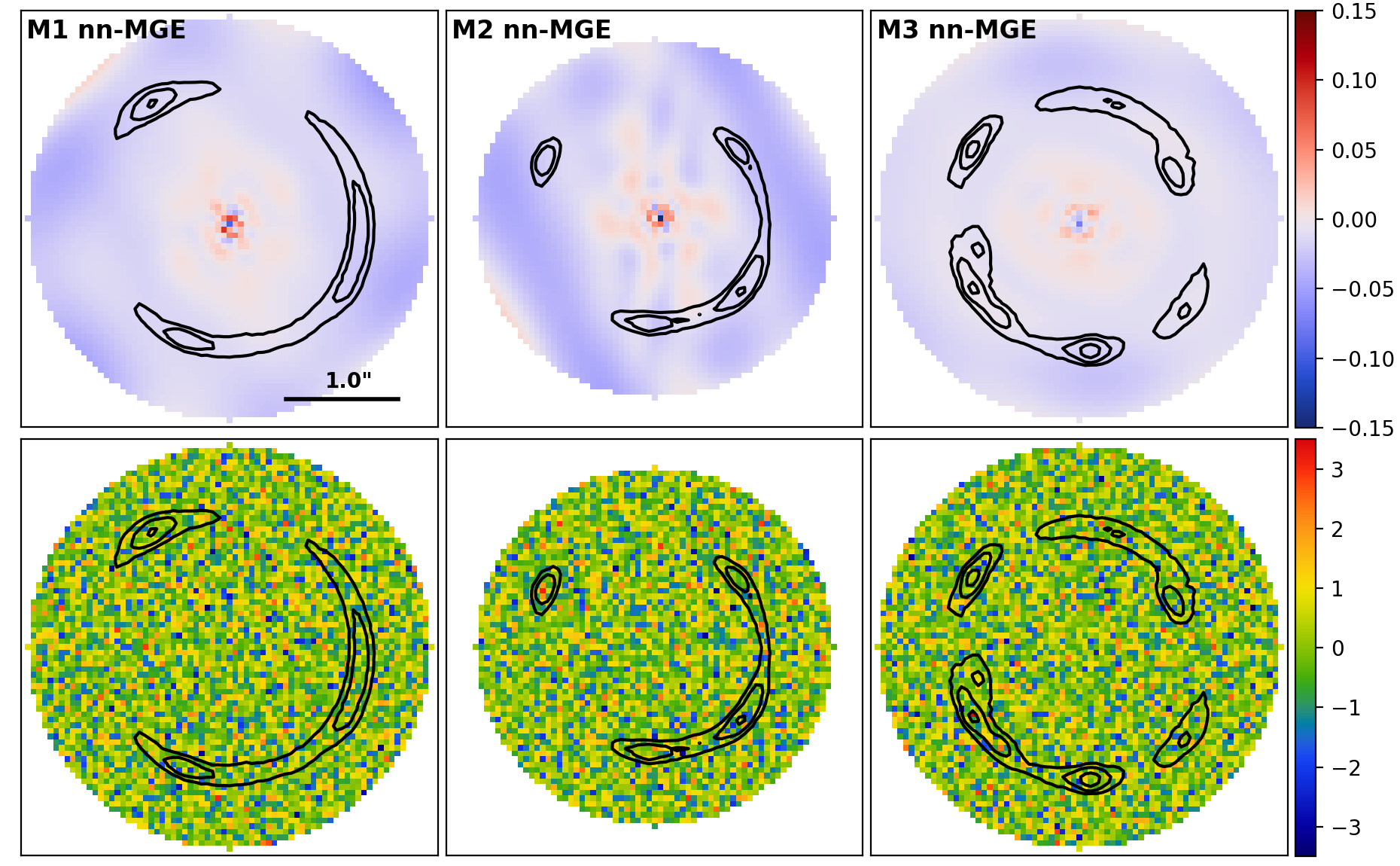}
    \caption{\textbf{Top row:} Relative differences between the best-fit ``nn-MGE'' lens light model and the input lens light. \textbf{Bottom row:} Normalized residuals of the best-fit lensing model. The black contours mark the source emission (image data - best-fit lens light model).}
    \label{fig:mock_results_nn_MGE}
\end{figure*}

\begin{figure*}
    \includegraphics[width=2.0\columnwidth]{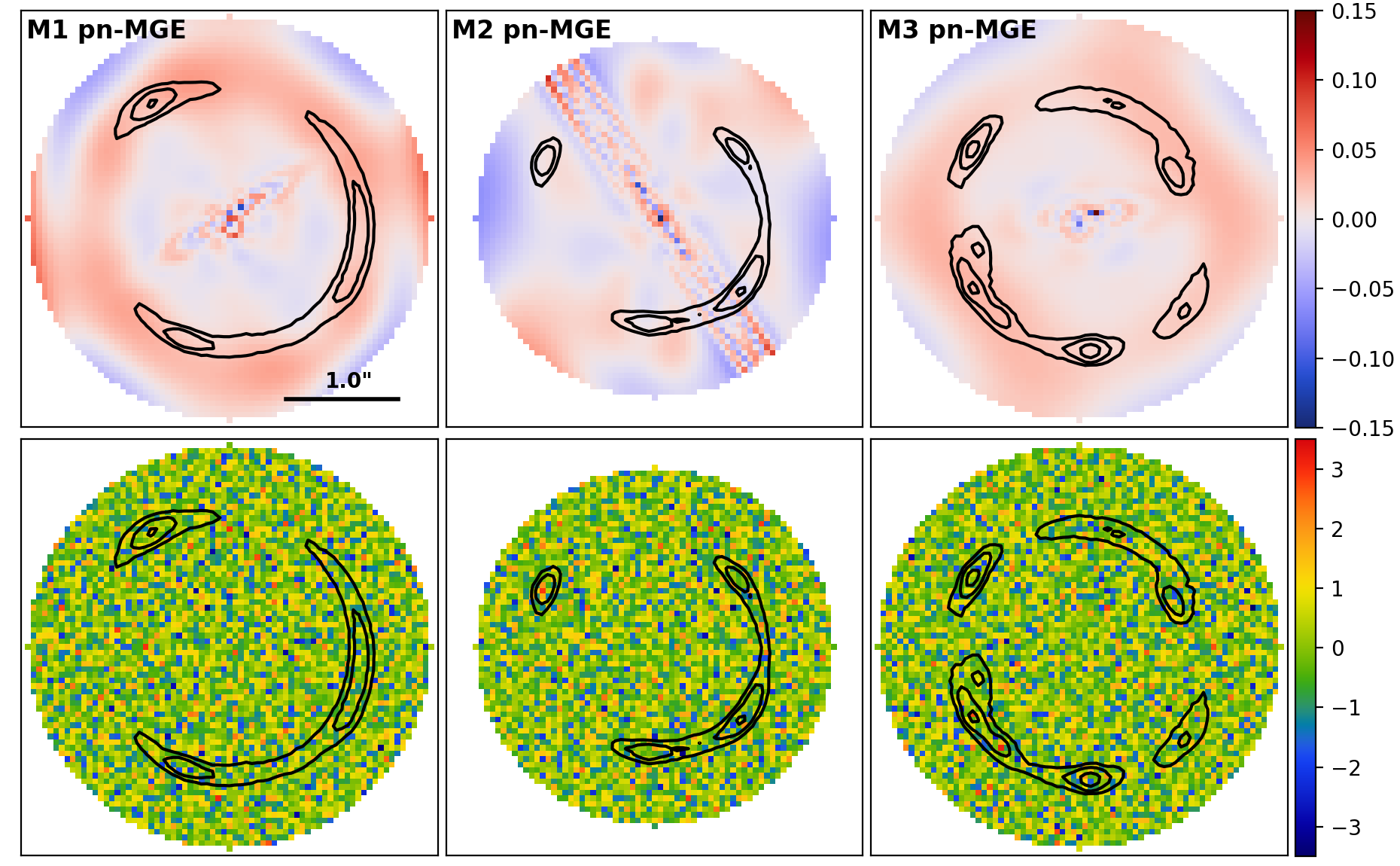}
    \caption{\textbf{Top row:} Relative differences between the best-fit ``pn-MGE'' lens light model and the input lens light. \textbf{Bottom row:} Normalized residuals of the best-fit lensing model. The black contours mark the source emission.}
    \label{fig:mock_results_pn_MGE}
\end{figure*}

\begin{figure*}
    \includegraphics[width=2.0\columnwidth]{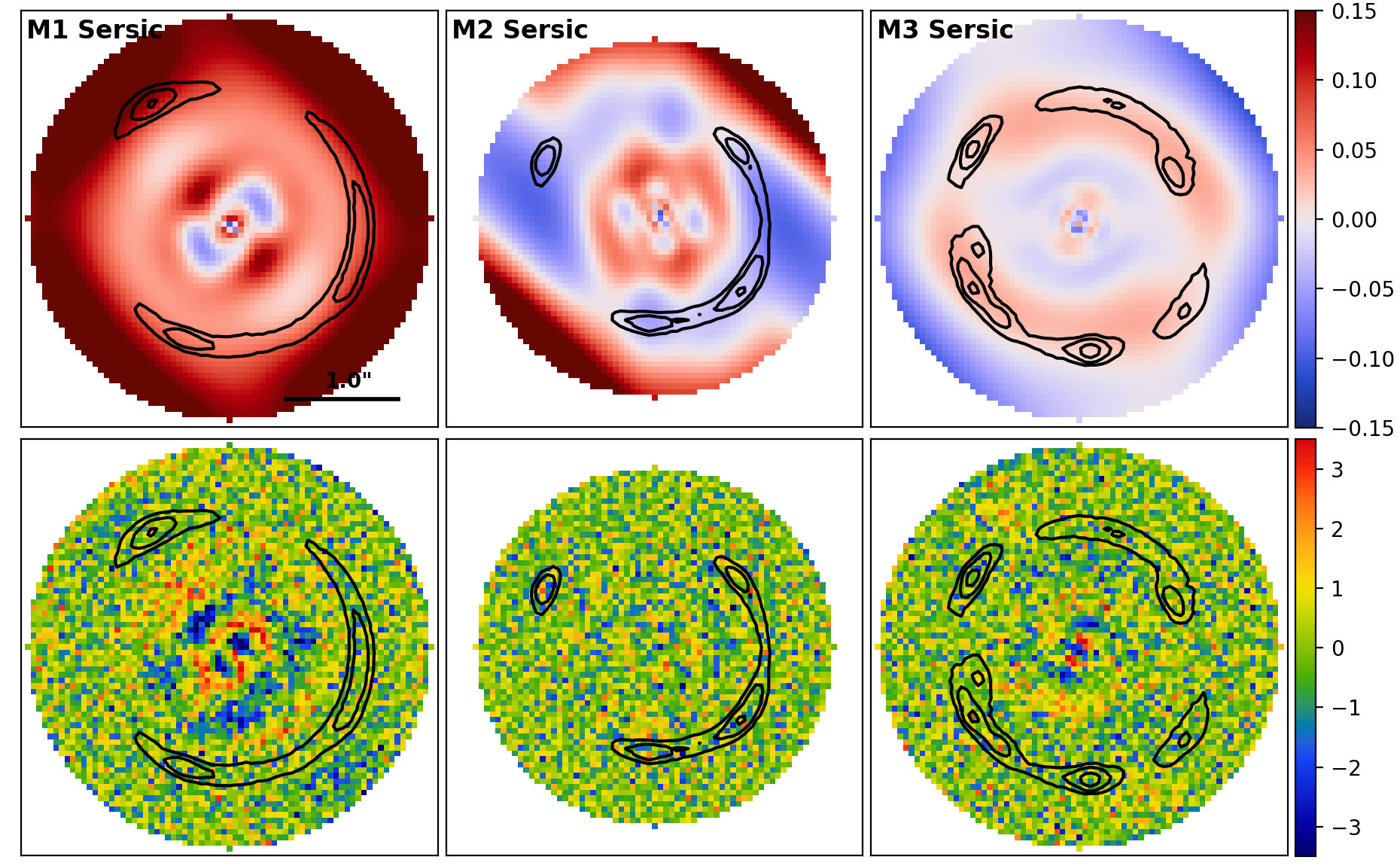}
    \caption{\textbf{Top row:} Relative differences between the best-fit 3 S\'ersics lens light model and the input lens light. \textbf{Bottom row}: Normalized residuals of the best-fit lensing model. The black contours mark the source emission.}
    \label{fig:mock_results_Sersic}
\end{figure*}


We show results using the ``nn-MGE'' model in Fig.~\ref{fig:mock_results_nn_MGE}. Different columns show the results for different mock data. The first row shows the ``lens light relative difference map'' ((model lens light - input lens light) / input lens light), therefore showing the relative difference between the best-fit MGE model and the true lens light input. The model recovers the true input lens light to a very high accuracy, with relative differences below 5\% everywhere except the central regions, which are due to Poisson noise. The second row shows the overall (including both lens and source emission) normalized residuals ((best-fit model image - data) / errors) for the fits with an MGE lens light model. The model fits the mock data to the noise level, exhibiting no significant correlated residuals.  Table~\ref{tab:mock_results} lists the best-fit parameters of our power-law mass model, with values in brackets indicating the $3\sigma$ confidence regions. The model correctly recovers all of the parameters of the input power-law mass profiles within the $3\sigma$ error.

In Fig.~\ref{fig:mock_results_pn_MGE}, we show the same plots for fits using the ``pn-MGE'' model. The lens light relative difference map shown in the top row shows systematic issues for all three mock lenses. Undesirable ``ring-like'' correlated residuals around the Einstein arc of M1 and M3 are seen, meaning that the ``pn-MGE'' model is systematically overestimating the lens light fluxes in areas overlapping with the source emission. The lens light relative difference map of M2 shows non-physical stripes that alternate between positive and negative values. Inspection of the MGE intensities reveals Gaussian profiles with alternating positive and negative intensities that counterbalance each other, indicating overfitting. The bottom row shows this model again fits the mock data to the noise level. The second part of Table~\ref{tab:mock_results} lists the best-fit parameters of the power-law mass model when using the ``pn-MGE'' lens light model. Despite the issues seen in the lens light relative difference map, this model still correctly recovers the input power-law mass profiles.
 
\begin{table*}
    \centering
    \begin{tabular}{ccccccc} 
    \hline
    & $x$~[$\arcsec$] & $y$~[$\arcsec$] & $R_{\rm E}~$[$\arcsec$] & $q$ & $\phi$~[$^\circ$] & $\gamma$ \\
    \hline
    M1 nn-MGE & 0.014~(0.010, 0.018) & -0.068~(-0.072, -0.065) & 1.198~(1.178, 1.218) & 0.80~(0.77, 0.83) & 45.6~(43.1, 47.9) & 2.08~(1.92, 2.23) \\
    M2 nn-MGE & 0.047~(0.044, 0.051) & 0.017~(0.014, 0.019) & 1.006~(0.990, 1.023) & 0.77~(0.74, 0.80) & 59.6~(58.0, 61.1) & 1.94~(1.83, 2.05) \\
    M3 nn-MGE & -0.003~(-0.005, 0.000) & -0.033~(-0.035, -0.031) & 1.101~(1.088, 1.124) & 0.75~(0.72, 0.77) & 31.2~(30.2, 32.5) & 2.01~(1.94, 2.15) \\
    \hline
    M1 pn-MGE & 0.015~(0.010, 0.018) & -0.068~(-0.072, -0.065) & 1.199~(1.178, 1.222) & 0.80~(0.76, 0.82) & 46.1~(43.4, 47.6) & 2.08~(1.91, 2.25) \\
    M2 pn-MGE & 0.047~(0.044, 0.051) & 0.015~(0.012, 0.018) & 1.013~(0.995, 1.032) & 0.76~(0.73, 0.79) & 59.2~(57.3, 60.6) & 1.99~(1.86, 2.13) \\
    M3 pn-MGE & -0.000~(-0.004, 0.003) & -0.030~(-0.032, -0.027) & 1.108~(1.089, 1.136) & 0.74~(0.70, 0.77) & 32.6~(29.1, 33.7) & 2.04~(1.92, 2.20) \\    
    \hline
    M1 $3\times$S\'ersic & 0.013~(0.009, 0.017) & -0.067~(-0.072, -0.064) & 1.184~(1.177, 1.193) & 0.80~(0.78, 0.82) & 46.0~(43.6, 47.5) & 1.96~(1.91, 2.02) \\
    M2 $3\times$S\'ersic & 0.026~(0.022, 0.035) & 0.010~(0.006, 0.023) & 1.05~(0.99, 1.09) & 0.71~(0.68, 0.98) & -4.2~(-17.7, 9.5) & 2.72~(2.61, 2.82) \\
    M3 $3\times$S\'ersic & -0.006~(-0.009, -0.003) & -0.032~(-0.034, -0.030) & 1.14~(1.12, 1.15) & 0.73~(0.69, 0.74) & 32.1~(30.1, 33.4) & 2.27~(2.18, 2.36) \\
    \hline
    \end{tabular}
    \caption{Best-fit power-law mass model parameters using MGE / S\'ersic lens light model. Values in brackets indicate 3$\sigma$ confidence regions.}\label{tab:mock_results} 
\end{table*}

Finally, we show the results using the S\'ersic lens light model in Fig.~\ref{fig:mock_results_Sersic}, following the same format as Fig.~\ref{fig:mock_results_nn_MGE} and Fig.~\ref{fig:mock_results_pn_MGE}. The first row shows the relative differences between the best-fit 3~$\times$~S\'ersic lens light model and the input lens light. The differences are much larger than seen for either MGE model shown before, particularly for M1 and M2, where in some regions the relative differences can exceed 15\%. The second row shows the normalized residuals, where the 3~$\times$~S\'ersic lens light model produces a poor fit to the image with significantly correlated residuals. The best-fit power law parameters are listed in Table~\ref{tab:mock_results}, where a biased estimation of the lens mass profile is seen. Specifically, the slopes are recovered incorrectly at $3~\sigma$ confidence for all three mocks. This discrepancy arises because the lens mass model compensates for inaccuracies in the lens light model by adjusting its location, mass, shape and other attributes to fit the residuals left by an imprecise representation of the lens light.

\begin{figure*}
    \includegraphics[width=2.0\columnwidth]{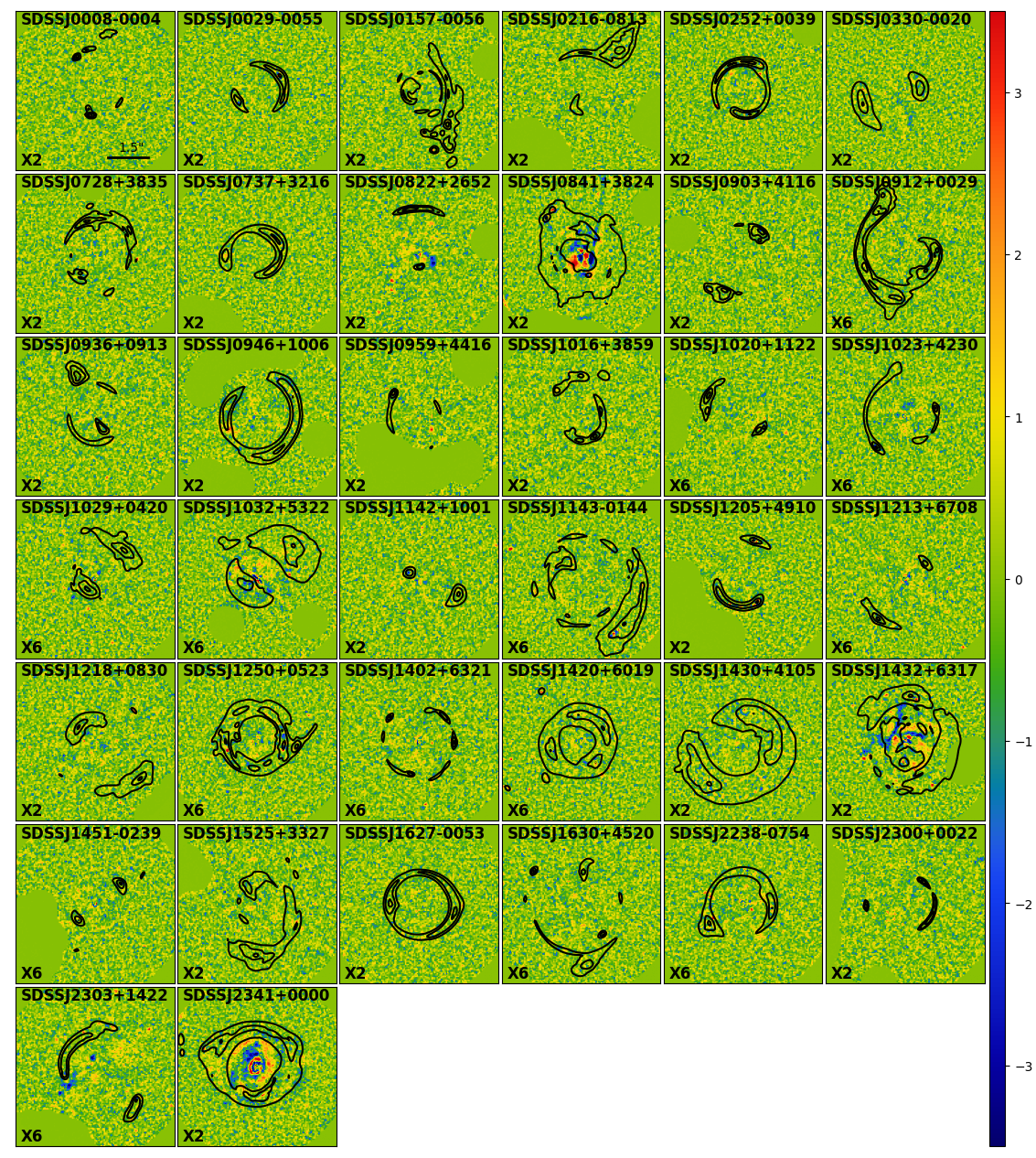}
    \caption{Best-fit normalized residuals for images of 38 SLACS lenses. The letter in the bottom left corner of every panel indicates the lens light model we used to fit the image. ``X2'' means a ``$2\times30 + 1\times10$'' MGE lens light model while ``X6'' means a ``$6\times30 + 1\times10$'' MGE lens light model. The black contours mark the source emission.}
    \label{fig:slacs_results}
\end{figure*}

\begin{figure*}
    \includegraphics[width=2.0\columnwidth]{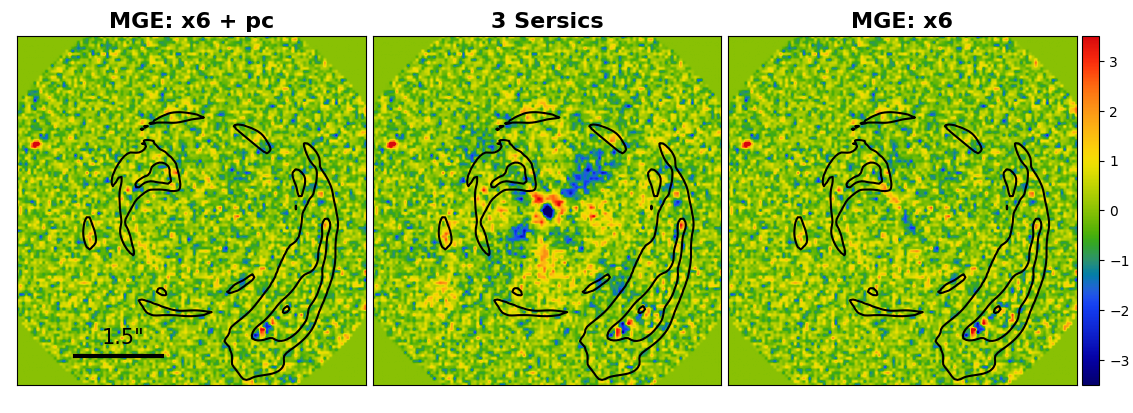}
    \caption{Normalized residuals for SDSSJ1143-0144 using three different lens light models, the ``$6\times30 + 10$'' MGE model (left panel), the 3 Sersic model, the ``$6\times30$'' MGE model. The colour bar range is from -3.5 to +3.5. The black contours are the same as those in Fig.~\ref{fig:slacs_results}.}
    \label{fig:example_norms}
\end{figure*}

\subsection{Summary}

The MGE models fit all three mock images with complex lens and source morphologies to the noise level. However, without enforcing non-negative constraints on the solutions of Gaussian intensities and source pixel fluxes, the ``pn-MGE'' model overestimates the lens light flux in regions where the lensed source is observed and may suffer from the over-fitting due to the Gaussians alternating between large positive and negative values. By enforcing non-negative constraints on the solution, the ``nn-MGE'' fixes these issues and accurately recovers the input lens light distribution with a $>95$\% accuracy. It also accurately recovers the lens mass profile parameters. The MGE is a significant improvement on a standard 3~$\times$~S\'ersic lens light model. The MGE lens light model demonstrates the ability to capture the surface brightness of a realistic lens galaxy.

\section{Application to SLACS lenses}\label{sec:slacs_results}

\subsection{Fitting results}
\subsubsection{MGE fitting results}
We now apply the MGE lens light model with the non-negative constraints to 38 SLACS lenses. We first fit a ``$2\times30 + 1\times10$'' MGE model (2 sets of 30 Gaussians plus an additional set of 10 Gaussian profiles which capture central point source emission). However, in certain SLACS lenses the model left lens light residuals and we therefore additionally fitted a more complex ``$6\times30 + 1\times10$'' MGE model (6 sets of 30 Gaussians with an additional point-source set). We fit the lens mass using an elliptical power law profile plus an external shear and use a Voronoi mesh for the source. Some SLACS lenses have several nearby galaxies within their $3.0\arcsec$ circular mask. To avoid light contamination from those nearby galaxies, following the same procedure of \citet{Nightingale2024}, we have manually selected out the region of contamination and increased the related noise by several orders to ensure the light of nearby galaxies would not affect our analysis. 
\begin{figure*}
    \includegraphics[width=2.0\columnwidth]{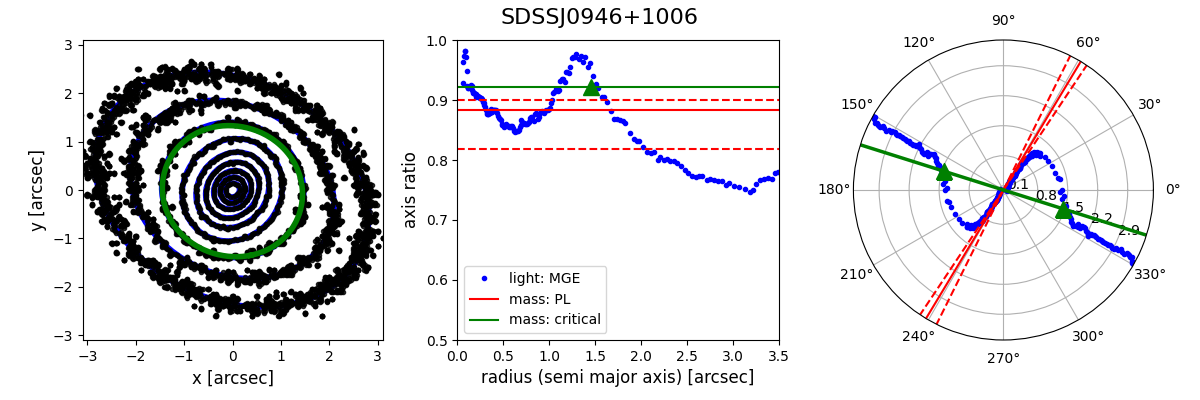}
    \caption{\textbf{Left panel:} Black lines represent the contours of the best-fit MGE lens light model. Blue curves are the best-fit ellipses to the contours. The green curve marks the critical line of the best-fit power-law mass profile. \textbf{Middle panel:} Blue points show the axis ratios of the isphotes of the best-fit MGE lens light model as a function of radius. The horizontal green line indicates the axis ratio of the critical curve of the best-fit profile. The green triangle marks the Einstein radius. The horizontal red line represents the axis ratio of the best-fit power-law profile with dashed lines showing its associated $3\sigma$ errors. \textbf{Right panel}: Blue points show the position angles of the isophotes of the best-fit MGE lens light model as a function of radius. The green line indicates the position angle of the critical curve of the best-fit mass profile. The green triangle marks the Einstein radius. The red line represents the position angle of the best-fit power-law mass profile, with dashed lines showing its $3\sigma$ errors.}
    \label{fig:isophotes_example}
\end{figure*}

Fig.~\ref{fig:slacs_results} shows the normalized residuals of our fits to the 38 SLACS lensing systems, with the colour bar ranging from $-3.5\sigma$ to $+3.5\sigma$. The bottom-left corner of each panel shows the MGE lens light model used for fitting the image. ``X2'' denotes a ``$2\times30 + 1\times10$'' MGE model, while ``X6'' denotes a ``$6\times30 + 1\times10$'' MGE model. Black contours outline the morphology of source emission. The MGE lens light model fits the majority of SLACS lenses to the noise level. However, for three lenses---SDSSJ0841+3825, SDSSJ1432+6317 and SDSSJ2341+0000, significant (> 3$\sigma$) correlated normalized residuals are observed in regions far from the source emission. These residuals are due to an inaccurate lens light model, caused by distinct asymmetric features that even an MGE model cannot fit, such as the ``spiral''-like feature shown in SDSSJ1432+6317.

\subsubsection{Three S\'ersic lens light model results}

For comparison, we have also applied the 3 S\'ersic lens light model to analyze the SLACS lenses. The normalized residuals of this analysis are shown in Fig.~\ref{fig:slacs_results_3Sersics}. The model fits the SLACS images significantly worse than the MGE light model, with around half of the fits exhibiting significant correlated residuals related to the lens light. Taking SDSSJ1143-0144 as a case study, the left and middle panels of Fig.~\ref{fig:example_norms} compare the normalized residuals of the ``$6\times30 + 1\times10$'' MGE lens light model and the 3 S\'ersic lens light model. The three S\'ersics model shows significant correlated residuals and therefore fails to capture angular complexity inherent in the light distribution of the lens galaxy, which the MGE model fits successfully. While one could theoretically increase the number of S\'ersic profiles to account for these residuals, this becomes impractical as it significantly increases the complexity of the parameter space due to the addition of more parameters. 


\subsubsection{``Point centre'' MGE set}\label{sec:point_centre}
Our MGE lens light model also includes a point source of light, consisting of a set of 10 Gaussians with a centre distinct from the other MGE sets. This fits point source emission observed in the very central region of a subset of SLACS lenses. Fig.~\ref{fig:slacs_results_MGE_nopoint} shows MGE fits to the SLACS sample without including this set. In approximately one-quarter of the cases removing the set of 10 Gaussians leads to point-like residuals appearing at the centre of the lens galaxy, illustrating that it serves an important role in capturing this central point-like emission accurately. Fig.~\ref{fig:example_norms} compares the normalized residuals for SDSSJ1143-0144 using an MGE lens light model with (left panel) and without (right panel) this set of 10 Gaussians. Omitting them leads to the presence of minor dipole residuals in the central region of the lens galaxy's light distribution.

\subsection{Complexity of realistic lens light emission}
\subsubsection{Ellipticity and position angles}
For 35 out of 38 SLACS lenses with satisfactory MGE fits, we assess the complexity of lens light emission based on the changing ellipticity and position angles of their isophotes as a function of the 2D radius. For example, in Fig.~\ref{fig:isophotes_example}, we show isophote analysis of the best-fit MGE lens light models of SDSSJ0946+1006. To make the visuals close to observations, we incorporate the PSF and realistic noise into the best-fit lens light model for this analysis. In the left panel, black contours represent the isophotes of the best-fit MGE lens light model. Each contour is fitted with an elliptical curve, depicted by blue lines in the plot. The green curve represents the critical curve on the lens plane. We measure the axis ratios and position angles of the best-fit elliptical curves. The blue points in the middle and right panels show the axis ratios and position angles, respectively, as functions of the (semi-major axis) radius. In the middle panel, the green horizontal line represents the axis ratio of the critical curve, with the green triangle marking the Einstein radius. The red horizontal line shows the axis ratio of the best-fit power-law profile, with dashed lines depicting its 3$\sigma$ errors. Similarly, in the right panel, the green line and triangles denote the position angle of the critical curve and the Einstein radius, respectively. The red solid and dashed lines represent the position angle of the best-fit power law profile and its associated $3\sigma$ errors. We can see that SDSSJ0946+1006 exhibits a highly complex surface brightness with isophotes twisting nearly $90^\circ$ from the inner to outer regions. (Similar plots as Fig.~\ref{fig:isophotes_example} for all 35 SLACS lenses can be found at \url{https://github.com/qiuhan96/isophotes_MGE_SLACS}.)

\begin{figure}\includegraphics[width=1.0\columnwidth]{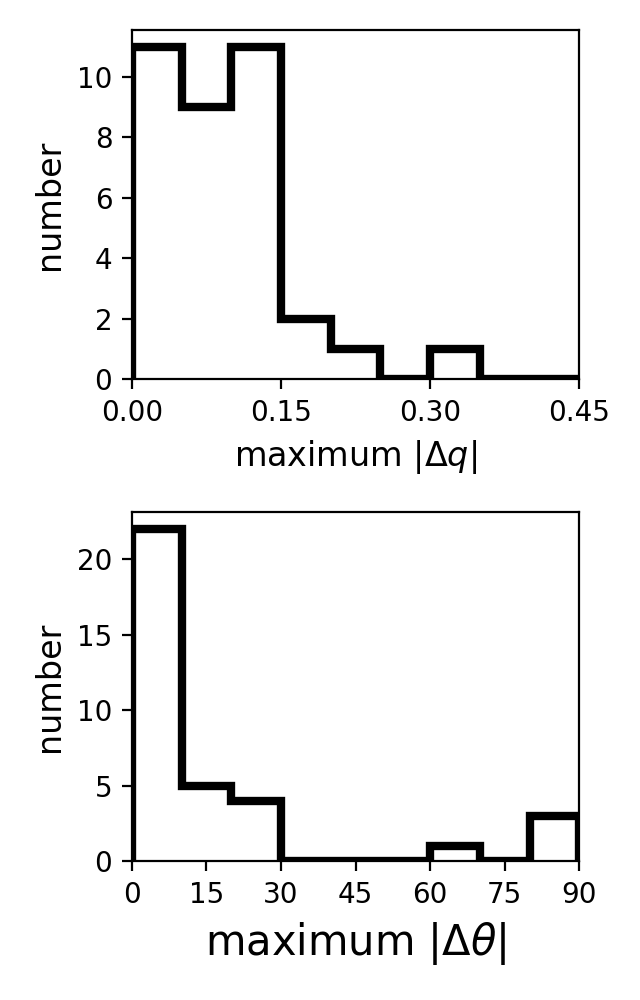}
    \caption{\textbf{Top panel:} The histogram of the maximum changes in axis ratios within the range of $0.5\arcsec$ to $2.5\arcsec$ for 35 SLACS lenses. \textbf{Bottom panel:} The histogram of the maximum changes in position angles for those lenses.}
    \label{fig:max_q_phi}
\end{figure}

For an overview of the complexity in the lens light emission for the 35 SLACS lenses, Fig.~\ref{fig:max_q_phi} summarizes the maximum changes in axis ratios and position angles within the range of $0.5\arcsec$ and $2.5\arcsec$, covering the Einstein radii and the majority of lens light emission. The top panel displays the histogram of the maximum change in axis ratios and the bottom panel shows the histogram of the maximum change in position angles. Notably, a significant portion of SLACS lenses ($\sim40\%$) exhibit isophotes with changes in axis ratios exceeding 0.1 or changes in position angles over $10^\circ$. This suggests that the phenomenon of twisting non-elliptical isophotes is common among typical lens galaxies.

\subsubsection{Boxiness~/~Diskiness}
By analysing the isophotes of the best-fit MGE light models we can also assess the boxiness~/~ diskiness of the 35 SLACS lenses. To quantify an isophote's shape, we follow the methodology outlined by \citet{Bender1987}, employing Fourier analysis to measure deviations from standard elliptical shapes. The equation for this analysis is as follows:
\begin{equation}
    \delta R\left(\theta\right) = R\left(\theta\right) - R_{\rm el}\left(\theta\right) = a_0 + \sum_{1}^{N}\left(a_n\cos{n\theta} + b_n\sin{n\theta}\right),
\end{equation}
where $R(\theta)$ represents the polar coordinate of a given isophote, and $R_{\rm el}(\theta)$ is the polar coordinate of the ellipse best-fitting the isophote. The deviation, $\delta R(\theta)$, is expressed in terms of Fourier coefficients $a_n$ and $b_n$ (for $n=0, 1, ..., N$), which capture the deviation's amplitude and phase. We use the ratio $a_4 / a$, where $a$ is the semi-major radius of the isophote, to quantify the isophote's shape. A positive value of $a_4/a$ indicates a ``disky'' isophote, whereas a negative value indicates a ``boxy'' isophote.

\begin{figure}\includegraphics[width=1.0\columnwidth]{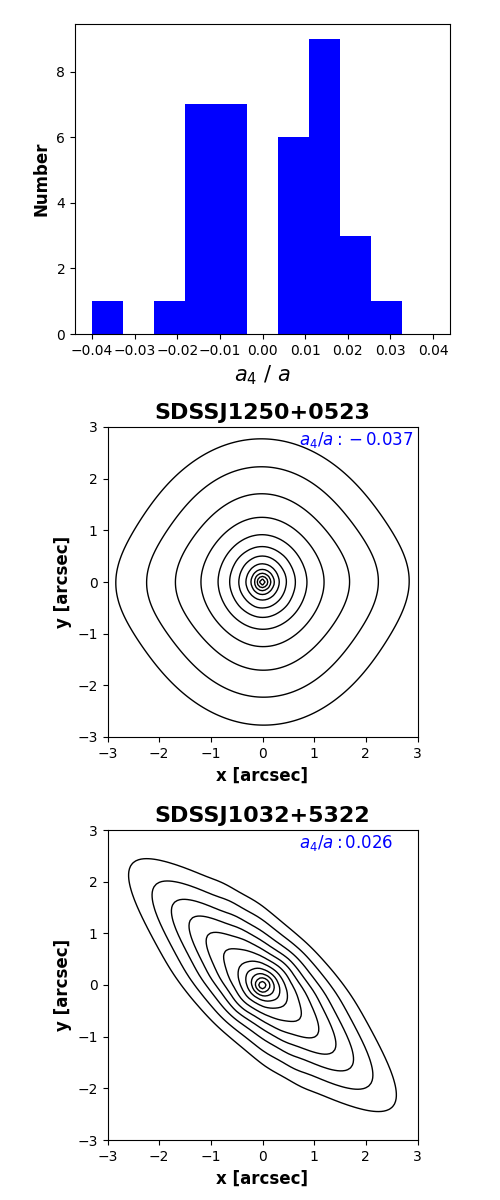}
    \caption{\textbf{Top panel:} The histogram of the $a_4 / a$ of maximum absolute values within the range of $0.5\arcsec$ to $2.5\arcsec$ for 35 SLACS lenses. \textbf{Middle panel:} The isophotes of the best-fit lens light model of SDSSJ1250+0523. This lens has the most boxy isophotes in our analysis. \textbf{Bottom panel:} The isophotes of the best-fit lens light model of SDSSJ1032+5322. This lens has the most disky isophotes in our analysis.}
    \label{fig:boxy}
\end{figure}

We report the $a_4 / a$ with the maximum absolute value within the range of $0.5\arcsec$ and $2.5\arcsec$ for each lens galaxy. The histogram presented in the top panel of Fig.~\ref{fig:boxy} shows the distribution of these maximum $|a4 / a|$ values. We find that the vast majority of the lenses, 32 out of 35, have a maximum $|a_4 / a|$ less than 0.02. Approximately half of the lens sample demonstrates a noticeable degree of boxiness~/~diskiness, with their maximum $|a_4 / a|$ falling within the range of 0.01 to 0.02. Among the analyzed systems, SDSSJ1250+0523 has the most boxy isophotes, characterized by an $a_4 / a$ value of $-0.037$, while SDSSJ1032+5322 shows the most disky isophotes, with an $a_4 / a$ value of 0.026. The middle and bottom panels of Fig.~\ref{fig:boxy} show the best-fit isophotes of these two lenses. 

From the analysis of the shape of isophotes, it is clear that while the majority of the SLACS lenses do not exhibit significant boxiness or diskiness, with the majority of lenses having a maximum $|a_4/a|$ smaller than 0.02, there are notable exceptions that exhibit distinct boxy or disky shapes, with $|a_4/a|$ exceeding 0.02. This diversity highlights the variation in structural features across different lens galaxies, with a small subset showing significant departures from elliptical symmetry.

\section{Discussion}\label{sec:discussion}

\subsection{Origin of lens light complexity}

The MGE model has demonstrated a remarkable capability to accurately fit the light distribution of most real lenses, uncovering complexities like radial ellipticity variations, twisting isophotes, and boxiness~/~diskiness. Similar features have been observed in S0 and elliptical galaxy populations \citep{Hao2006}, including detailed stellar kinematics studies \citep{ Emsellem2011, Krajnovic2011}. However, strong lens selection effects produce lens galaxies with higher stellar masses than these galaxy samples, which is correlated with these features. The MASSIVE survey studies about 100 of the most massive nearby elliptical galaxies, with comparable masses to SLACS lenses. We therefore compare our results to \citet{Goullaud2018}, who conducted a detailed photometric study of 35 massive ellipticals. Their methodology differs markedly from ours, therefore the purpose of this comparison is to simply confirm that the luminous complexity seen in SLACS lenses is reported in samples of analogous unlensed massive ellipticals. 

\citet{Goullaud2018} observe changes in galaxy ellipticities of $0.1 - 0.2$ over a radial range of $1 - 10$~kpc, comparable to the axis-ratio changes shown for SLACS lenses in Fig.~\ref{fig:max_q_phi}. $37\%$ of their sample show variations in the position angle exceeding 20 degrees, comparable to values shown in Fig.~\ref{fig:max_q_phi}. Boxy and disky isophotes are observed in their sample with magnitudes of $0.01 - 0.02$ and an approximately fifty-fifty ratio of boxy and disky, consistent with the values and distribution shown for SLACS in Fig.~\ref{fig:boxy}. The deviations of SLACS lenses from elliptical isophotes are therefore broadly consistent with nearby massive elliptical galaxies.

The isophotal analysis offers clues on a galaxy's formation history. Ellipticity variations and twists are readily explained by an underlying triaxial 3D mass distribution which manifests in 2D projection. Stellar kinematic analysis reveals massive ellipticals often show evidence for trixality, which is likely a product of their merger history and growth. Even in prolate or oblate systems, recent merger activity may produce departures from ellipticity. However, morphological structures like bars and the presence of dust in the centres of galaxies can also produce these features \citep{Goullaud2018}. Our MGE technique therefore has the potential to reveal more about the formation history of strong lens galaxies.

We also observe central point-like emission in the centre of $\sim 25\%$ of lens galaxies. Galaxies often exhibit central point-like emission (e.g. \citet{Bruce2016}), which can be due to faint emission associated with the galaxy's central supermassive black hole or nuclear stellar emission. These phenomena do not offer much information on the galaxy's formation history. However, a tantalizing alternative is that these could be a dim lensed source central image, made possible by the lens's mass being centrally cored \citep{Winn2004, Quinn2016}. Whilst a somewhat unlikely hypothesis, the MGE does allow us to separate this point-like emission from the main lens's emission, motivating future studies which fit cored mass profiles to SLACS lenses.

\subsection{Implications for the complexity of lens mass distribution}

Recent studies on galaxy-galaxy strong lensing have shown that a single elliptical power-law mass model is insufficient to accurately describe the lens mass distribution, especially for scientific cases involving subhalo detection and inference of $H_0$, where precise lens mass modelling is crucial \citep{Hsueh2017, Hsueh2018, Cao2021, Van2021, He2023, Nightingale2024}. 
Lens mass complexity, beyond the power-law model, primarily arises from the stellar component of the lens galaxy because it dominates the mass within $\sim10$~kpc, the scale of galaxy-galaxy strong lensing. Angular complexity is particularly important, because strong lensing is highly sensitive to the projected enclosed mass within the critical curve, and any variations in the 2D angular shape of the mass distribution can significantly alter the shape of the critical curve. 

Our analysis of the isophotes of 35 SLACS lenses confirms that it is common for a lens galaxy's surface brightness to exhibit angular complexity beyond a single ellipse, which a single elliptical power-law does not fully capture. This includes twisting (arising from the varying position angles) and boxiness or diskiness (arising from the varying ellipticity). Supposing that stellar mass roughly traces the stellar light, it is reasonable to suggest that the stellar mass would exhibit a similar level of complexity as the surface brightness. There are hints of this in the literature, for example, our study notes that SDSSJ1250+0523 is the lens with the most boxy isophotes, and the same system is flagged by \citet{Nightingale2024} to require an M4 multipole (which represents a boxy mass distribution) in order to produce a robust dark matter inference. \citet{Stacey2024} also find evidence that a mass-model multipole correlates with lens light emission. Angular complexity in the stellar mass could explain puzzling results regarding the external shear of strong lenses \citep{Gomer2020, Gomer2021, Etherington2023} and inconsistencies between power-law lensing and stellar dynamics measurements \citep{Etherington2023b}.

Our MGE lens light model is well suited to being used in a two-component decomposed mass model, where the MGE represents the stellar mass and is modelled simultaneously with a dark matter halo. The ease of computing deflection angles from Gaussian profiles facilitates this \citep{Shajib2019}. This model is more physically motivated than a power law and scientifically could open many doors, for example measuring the centre, mass and ellipticity of the dark halo hosting the lens galaxy, which are challenging to measure with alternative methods like stellar dynamics. Decomposed mass models have previously been used \citep{Nightingale2019, Schuldt2019, Chen2022, Nightingale2024}, however, these studies rely on light profiles like the S\'ersic model which we have shown in this work are limited in their applicability. The next steps are therefore for us to begin fitting MGE mass models and to investigate assumptions regarding the conversion of stellar light to stellar mass, for example, whether a constant mass-to-light ratio is sufficient.

\subsection{Automated lens modelling with an MGE}

With samples of over 100,000 strong lenses incoming \citep{Collett2015}, automated lensing analysis is crucial \citep{Etherington2022}. The MGE offers significant advantages over traditional S\'ersic profiles in lens modeling, for both the lens galaxy and source galaxy. First, it provides enhanced flexibility in capturing complex light distributions, effectively reducing residuals from foreground lens light subtraction which often lead one to infer incorrect mass and source models \citep{Etherington2022}. Second, parameterizing the MGE requires only 4 non-linear parameters (compared to 6 for a linearized S\'ersic profile), leading to a reduction in the overall number of non-linear parameters. Crucially, the MGE's parameterization does not have a non-linear parameter governing the galaxy's size or luminosity (which are solved for linearly), eliminating a significant degeneracy between the source galaxy's size and the lens mass model parameters, ensuring more reliable sampling and reducing the risk of incorrectly inferring local maxima during optimization. Approaches like the B-spline lens light subtraction are user-intensive, requiring that the subtraction is performed manually via the graphical user interface. These benefits will be expanded upon by Franca et al., who perform automated analysis of 21 lenses using an MGE for the lens and source galaxies.

\section{Conclusions}\label{sec:conclusions}

We present a novel lens light model for galaxy-galaxy strong lensing analysis. The model is built on a multi-Gaussian expansion (MGE) method which has been widely used in stellar dynamical analysis. The key idea is to represent the lens light as sets of Gaussian profiles and linearly solve the intensities of each individual Gaussian profile. To simultaneously fit both the MGE lens light model and a pixelized source model, we incorporate the MGE lens light model into the conventional semi-linear scheme for pixelized source modelling \citep{Warren2003}. We note that because we are fitting two free-form models simultaneously, there could be a degeneracy between the lens light and pixelized source model, particularly in the lensed arc regions where both models can fit the light. To mitigate degeneracy between the MGE lens light model and the pixelized source model, we also introduce non-negative boundary constraints to the semi-linear inversion scheme which applies the physically motivated constraints that all lens and source emission must be non-negative. Compared to traditional lens light models like S\'ersic profiles, the primary advantage of an MGE lens light model is its ability to capture complex angular and radial features of real lenses using only a few parameters. 

To verify that the MGE lens light model can accurately fit realistic lens galaxies, we simulate three mock lensing images with realistic lens and source emission and employ them to test our MGE lens light model. The lens light is derived from three real SLACS lenses, while the source emission is represented by HST images of three nearby galaxies. We have demonstrated that the scheme utilizing the MGE lens light model can fit the mock images to the noise level (the second rows of Fig.~\ref{fig:mock_results_nn_MGE} and Fig.~\ref{fig:mock_results_pn_MGE}). For the MGE model with the non-negativity constraint, the relative difference between the best-fit lens light model and the true input is small, remaining below 5\% (the first row of Fig.~\ref{fig:mock_results_nn_MGE}). This result confirms that the MGE accurately captures the lens emission and that it enables a clean deblending of lens and source light. This is not the case when the non-negative constraint is relaxed (a common assumption in many lensing methods), with fits to the three mocks showing correlated residuals and evidence of over-fitting, because the MGE unphysically fits lensed source emission (the first row of Fig.~\ref{fig:mock_results_pn_MGE}). Fits using three S\'ersic profiles produce significant residuals (Fig.~\ref{fig:mock_results_Sersic}) and biased best-fit lens mass profiles (Table~\ref{tab:mock_results}).

Having confirmed that the MGE lens light model with the non-negative constraints enables a robust fit of a lens galaxy's emission, we apply it to 38 SLACS lenses. We judge that the MGE lens light model fits the majority -- 35 out of 38 of SLACS lenses -- accurately. We examine the ellipticity and position angles of the isophotes of these lenses. For approximately 40\% of the lenses, the axis ratio (position angle) of the isophotes can change as much as 0.1 ($10^\circ$) from $r=0.5\arcsec$ to $r=2.5\arcsec$ (Fig.~\ref{fig:max_q_phi}), suggesting that it is common for a lens galaxy to exhibit twisting non-elliptical stellar emission. We also measure the boxiness~/~diskiness of the isophotes using Fourier analysis to evaluate deviations from standard ellipses (Fig.~\ref{fig:boxy}). The majority of the 35 SLACS exhibit boxy or disky isophotes that are relatively small in overall magnitude, but large enough to impact lensing studies \citep{Cao2021, Van2020, Cohen2024}. We compare our results to a sample of unlensed massive elliptical galaxies with similar stellar masses to SLACS \citep{Goullaud2018} and confirm that the departures from ellipticity we have seen in SLACS lenses are observed with a similar frequency in non-lens samples. Approximately $\sim 25\%$ of lenses also require us to include an extra set of Gaussians which capture point-source emission at their centre, indicative of nuclear emission from the lens's black hole or supernova or possibly a central image of the background source.

Our study confirms that the stellar emission of nearly all strong lenses shows angular complexity which the model commonly used to model their mass -- the single elliptical power law -- can not capture accurately. The MGE model will therefore form the basis of a mass model which decomposes each lens galaxy into its stellar mass and dark matter. Future work seeks to fit this model to SLACS lenses and use it to inform studies of dark matter, cosmology and galaxy evolution.


{\color{blue}

}

\section*{Software Citations}

This work uses the following software packages:

\begin{itemize}

\item
\href{https://github.com/astropy/astropy}{{Astropy}}
\citep{astropy1, astropy2}

\item
\href{https://bitbucket.org/bdiemer/colossus/src/master/}{{Colossus}}
\citep{colossus}

\item
\href{https://github.com/dfm/corner.py}{{Corner.py}}
\citep{corner}

\item
\href{https://github.com/joshspeagle/dynesty}{{Dynesty}}
\citep{dynesty}

\item
\href{https://github.com/matplotlib/matplotlib}{{Matplotlib}}
\citep{matplotlib}

\item
\href{numba` https://github.com/numba/numba}{{Numba}}
\citep{numba}

\item
\href{https://github.com/numpy/numpy}{{NumPy}}
\citep{numpy}

\item
\href{https://github.com/rhayes777/PyAutoFit}{{PyAutoFit}}
\citep{Nightingale2021}

\item
\href{https://github.com/Jammy2211/PyAutoGalaxy}{{PyAutoGalaxy}}
\citep{pyautogalaxy}

\item
\href{https://github.com/Jammy2211/PyAutoLens}{{PyAutoLens}}
\citep{Nightingale2018, Nightingale2019, Nightingale2021, Nightingale2024}

\item
\href{https://www.python.org/}{{Python}}
\citep{python}

\item
\href{https://github.com/scikit-image/scikit-image}{{Scikit-image}}
\citep{scikit-image}

\item
\href{https://github.com/scikit-learn/scikit-learn}{{Scikit-learn}}
\citep{scikit-learn}

\item
\href{https://github.com/scipy/scipy}{{Scipy}}
\citep{scipy}

\end{itemize}

\section*{Acknowledgements}

QH, AA, CSF and SMC acknowledge support from the European Research Council (ERC) Advanced Investigator grant DMIDAS (GA 786910), to C.S.\ Frenk. JN is supported by an STFC/UKRI Ernest Rutherford Fellowship, Project Reference: RES/0583/7224. JPCF acknowledges CNPq (grant no. 140210/2021-0).

This paper makes use of both the DiRAC Data-Centric system and the Cambridge Service for Data-Driven Discovery (CSD3), project code dp004 and dp195, which are operated by the Institute for Computational Cosmology at Durham University and the University of Cambridge on behalf of the STFC DiRAC HPC Facility (www.dirac.ac.uk). These were funded by BIS National E-infrastructure capital grant ST/K00042X/1, STFC capital grants ST/H008519/1, ST/K00087X/1, ST/P002307/1, ST/R002425/1, STFC DiRAC Operations grant ST/K003267/1, and Durham University. DiRAC is part of the National E-Infrastructure.

\section*{Data Availability}
The data underlying this article will be shared on reasonable request to the corresponding author.



\bibliographystyle{mnras}
\bibliography{example} 




\appendix

\section{Natural Neighbour Pixelization and Cross-like Regularization}\label{appxA}

\begin{figure*}
    \includegraphics[width=2.0\columnwidth]{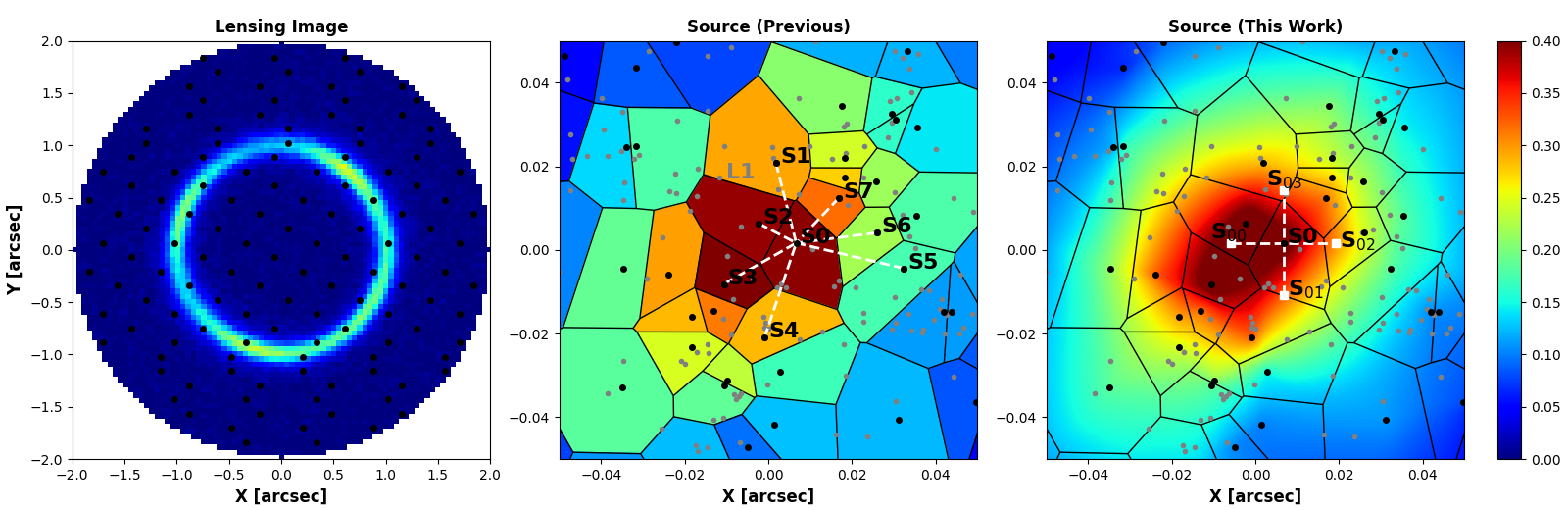}
    \caption{\textbf{Left panel:} An example  lensed mock image. Black points are the source grid points mapped to the image plane. \textbf{Middle panel:} Voronoi pixelization of \citet{Nightingale2015}. Black points mark the source pixel positions on the source plane. Grey points are image pixels traced to the source plane. Colours show the reconstructed fluxes of the source pixels. \textbf{Right panel:} New pixelization scheme used in this work, involving natural neighbour interpolation plus cross-like regularization.}
    \label{fig:voronoi_cells}
\end{figure*}

In Section~\ref{sec:source_pixelization}, we have described the semi-linear inversion framework for a pixelized source model. Here, we introduce the specific natural neighbour source pixelization and cross-like regularization scheme utilized in this work. We then show how it achieves a smooth likelihood curve and a robust estimation of lens parameters.

\subsection{Method}

\subsubsection{Mesh Centres}

The source pixelization first determines the centres of the Voronoi source pixels. Initial fits overlay a rectangular Cartesian grid of shape $(y_{\rm pix}, x_{\rm pix})$ over the image plane, which extends to and from the mask edges. All coordinates on this uniform grid which fall within the mask are retained and traced to the source plane via the mass model (pixels outside the mask are discarded). These coordinates, $M_i$, are used as the centre of the Voronoi cells, which therefore trace the mass model magnification.

Subsequent fits adapt the mesh centres $M_i$ to the source's unlensed morphology. This uses a previous model of the lensed source emission, $\Xi_{j}$, which is used to compute the weights
\begin{equation}
\label{eqn:KMeans}
W_j = (\frac{\Xi_{j} - \min{\Xi}} {\max{\Xi_{j}} - \min{\Xi}}) + W_{\rm floor} + \max{\Xi_{j}}. 
\end{equation}
The first term on the right-hand side runs from zero to one, where values closer to one correspond to the lensed source's brightest pixels. $W_{\rm floor}$ controls how much weight is given to the source's brightest pixels and is a free parameter in certain fits. $W$ is passed to a weighted k-means clustering algorithm \citep{scikit-learn} to determine image-plane coordinates which are traced to the source plane. The k-means assumes $N_{\rm pix}=1500$ source pixels throughout this work. This scheme adapts to the lensed source emission.

\subsubsection{Mapping Matrix}

The reconstruction computes the linear superposition of PSF-smeared source pixel images which best fits the observed image. This uses the mapping matrix $f_{ij}$, which maps the $j$-th pixel of each lensed image to each source pixel $i$, giving a total of $J$ lensed image pixels and $I$ source pixels. When constructing $f_{ij}$ we apply image-plane subgridding of degree $4 \times 4$, meaning that $16 \times J$ sub-pixels are fractionally mapped to source pixels with a weighting of $\frac{1}{16}$, removing aliasing effects \citep{Nightingale2015}. 

Each image sub-pixel is mapped to multiple Voronoi source pixels weighted via interpolation. We use Voronoi natural neighbour interpolation via Sibson's technique \citep{Sibson1981}. For every sub-pixel, $j$, the method considers a new polygon that adding this point to the Voronoi mesh computed from $M_i$ would create. The new polygon captures some of the area that was previously covered by its neighbours, which the method computes and uses to compute the interpolation weights in $f_{ij}$ as
\begin{equation}
\label{eqn:NaturalNeighbor}
w = f_{ij} = \frac{1}{\sum_{i=1}^{K} A_{\rm capture}} \sum_{k=1}^{K} A_{\rm capture} \, z_{k} \, ,
\end{equation}
where $K$ is the number of neighbours of a given Voronoi cell $i$ \footnote{More details about the natural neighbour interpolation technique can be found at \url{https://gwlucastrig.github.io/TinfourDocs/NaturalNeighborTinfourAlgorithm/index.html}.}. 

\subsubsection{Regularization}

Performing an inversion using \cref{eqn:NaturalNeighbor} by itself is ill-posed, therefore to avoid over-fitting noise the solution is regularized using a linear regularization matrix $H$ described by \citet{Warren2003}. The matrix $H$ applies a prior on the source reconstruction, penalizing solutions where the difference in the reconstructed flux of neighbouring Voronoi source pixels is large. Initial fits use gradient regularization (see \citet{Warren2003}) adapted to a Voronoi mesh (see \citealt{Nightingale2015}). The main results and those illustrated in this appendix use a scheme which adapts the degree of smoothing to the reconstructed source's luminous emission and interpolates values at a cross of surrounding points. The formalism for the calculation of these regularization matrices $H$ is given in Appendix A of \citep{Nightingale2024}.

\subsubsection{Inversion}

Following the formalism of \citet{Warren2003}, we define the data vector $\vec{D}_{i} = \sum_{j=1}^{J}f_{ij}(d_{j} - b_{j})/(\sigma_{j})^2$ and curvature matrix $F_{ik} = \sum_{j=1}^{J}f_{ij}f_{kj}/(\sigma_{j})^2$, where $d_{j}$ are the observed image flux values and $ b_{j}$ are the model lens light values. The source pixel surface brightnesses are given by $s = [F + H]^{-1} \vec{D}$ which are solved via a linear inversion that minimizes
\begin{equation}
\label{eqn:ChiSquared}
\chi^2 + G_{\rm L} = \sum_{j=1}^{J} \bigg[ \frac{(\sum_{\rm  i=1}^{I} s_{i} f_{ij}) + b_{j} - d_{j}}{\sigma_{j} } \bigg]^2 + s^{T}Hs \, .
\end{equation}
The term $\sum_{i=1}^{I} s_{i} f_{ij}$ maps the reconstructed source back to the image plane for comparison with the observed data and $G_{\rm L} = s^{T}Hs$ is a regularization term.

The degree of smoothing is chosen objectively using the Bayesian formalism given by Eq.~(\ref{equ: bayesian_evidence_MGE}).

\subsection{Likelihood Surface Noise}

In strong lensing analysis, modelling with pixelized source models may lead to noisy likelihood surfaces. This implies that a slight change in the parameter space of the lens mass could result in a disproportionately large jump in the likelihood \citep[see Figure 3 of][]{Etherington2022}. The noisy behaviour of a likelihood curve makes it difficult for a non-linear sampler to efficiently sample the parameter space, leading to an underestimation of parameter errors. Even worse, in complex strong lensing analyses, such as searching for small dark matter subhaloes, obtaining a robust estimate for the subhalo's mass and location becomes difficult, because the non-linear sampler may easily get trapped in a random local maximum due to a noisy likelihood surface.

The key reason for a noisy likelihood curve is the discreteness inherent in the construction of a pixelized source model. Taking the Voronoi source pixelization scheme of \citet{Nightingale2015} as an example, we illustrate the lensed image and the corresponding source reconstruction in Fig.~\ref{fig:voronoi_cells}. The left and middle panels depict the lensed image and the source reconstruction, respectively. Given the lens mass profile and the source grids on the lens plane, represented by black points in the left panel (1/5 of them shown for clarity), we can construct Voronoi cells on the source plane, which are the irregular cells in the middle panel, with black points indicating the locations of the source grids on the source plane. As depicted in the Voronoi tessellation, the source plane is naturally divided into regions corresponding to different source pixels. To establish a connection between the image pixels and the source pixels, the image pixels are also traced back to the source plane, marked as grey points in the middle panel (only 1/200 of them are shown for clarity). The image pixel is then assigned the same flux as the Voronoi cell to which the traced image pixels fall. With small changes of the lens mass parameter, such as the slope or the Einstein radius, the positions of points (both grey and black points) on the source plane will undergo slight adjustments. As the change is small, the fluxes of most traced image pixels (grey points) do not change significantly. However, for a point located right between two Voronoi cells, such as the point \textbf{L1} between the cell of \textbf{S1} and \textbf{S2} in the middle panel, a slight change in its location could entirely alter which cell it belongs to, resulting in a discontinuous change in its flux and, consequently, an unsmooth likelihood curve.

To alleviate this issue, we could introduce an interpolation scheme on the source plane so that we can ensure that when grey points smoothly change their locations on the source plane, their fluxes also change smoothly. The most commonly used 2D interpolation is a bilinear interpolation implemented upon a Delaunay tessellation. However, we note that although the bilinear interpolation provides a continuous source plane, the first derivative of the interpolation is not continuous at the boundaries of Delaunay cells. This characteristic makes it not ideal for analyzing subhalo signals, as a subhalo's perturbation to the lensed image is sensitive to the first derivative of the surface brightness of the source galaxy \citep{Vegetti2009a}. For this reason, we opt for the natural neighbour interpolation scheme \citep{Sibson1981} to represent the source plane. For a given point on a Voronoi source plane, the natural neighbour interpolation computes the value of that point as an area-weighted sum of the fluxes of its ``neighbours''. These neighbours are determined by treating this point as a new member added to the Voronoi tessellation and weighting the neighbours according to the overlap of the new Voronoi cell with the original Voronoi cells.
Despite the complexity of the algorithm, by design, the first derivative of natural neighbour interpolation is continuous, except at the sample points \citep{Hiyoshi2000}. In the right panel of Fig.~\ref{fig:voronoi_cells}, we show the pixelized source scheme of this work where the values of traced image pixels (grey points) are computed through the natural neighbour interpolation\footnote{For high performance, we have taken 
a C version  of this natural neighbour interpolation algorithm from \url{https://github.com/sakov/nn-c} modified it and made
a Python wrapper.
}. 


The mapping relation between the image and source plane is not the only source of the discreteness. We note that the regularization between source pixels could also give rise to issues. Following \citet{Nightingale2015}, the regularization of a source pixel is defined as the differences between the fluxes of the pixel and its neighbours, determined by the Voronoi tessellation. For example, the regularization of the source pixel \textbf{S0} is computed as (half the absolute) differences between its flux and fluxes of its seven neighbours, \textbf{S1}, \textbf{S2}, ..., \textbf{S7}. However, it is noted cells \textbf{S0} and \textbf{S1} are barely connected. We can imagine that a slight change in the points' locations could cause \textbf{S0} and \textbf{S1} to no longer be neighbours, resulting in two terms being lost in the regularization term. This, in turn, leads to a sudden change to the overall likelihood (evidence). The purpose of regularizing source pixels by flux differences between Voronoi neighbours is to ensure a reasonably smooth solution for our pixelized (free-form) source model. However, we note that it is actually not necessary to define the flux differences relying on a neighbour relation given by the Voronoi tessellation. Instead, similar to the regularization scheme of \citet{Vegetti2009a}, we define the regularization of a source pixel as the sum of absolute flux differences between the pixel and four cross points surrounding the pixel. For example, as shown in the right panel of Fig.~\ref{fig:voronoi_cells}, the regularization of the source point \textbf{S0} is computed as the sum of absolute flux differences between it and its four cross ``neighbours'', \textbf{S1}, \textbf{S2}, \textbf{S3} and \textbf{S4}. The distance between the source pixel and one of its cross points is determined as the square root of half of the associated Voronoi cell's area. The values of those cross points are computed through the natural neighbour interpolation. With this definition, a sudden change in a source pixel's neighbour relation would not lead to a sudden addition or subtraction of terms in the regularization. Instead, the total number of terms in the regularization remains fixed at four times the number of source pixels.     

\begin{figure}
    \includegraphics[width=1.0\columnwidth]{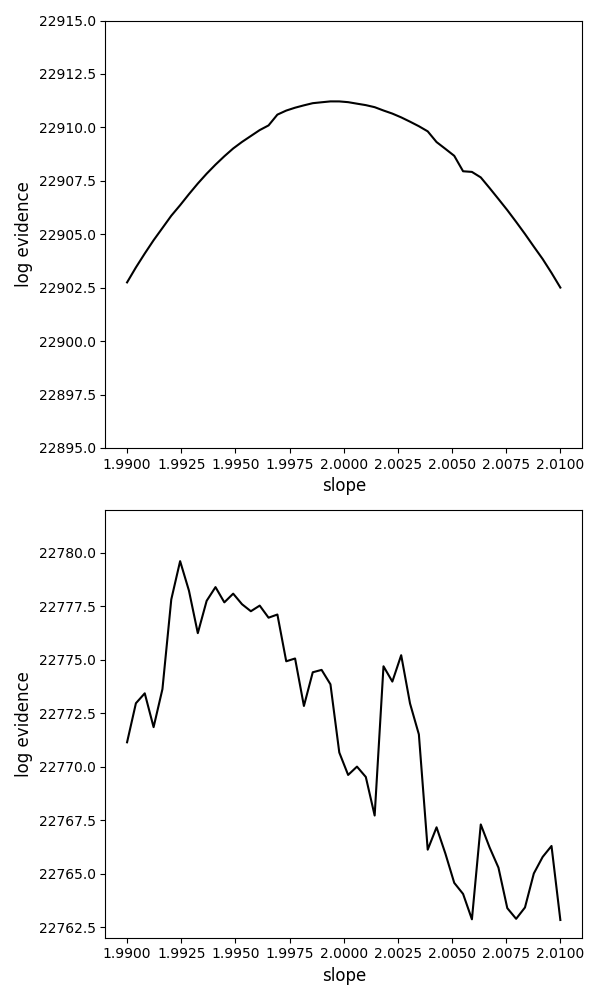}
    \caption{\textbf{Top panel:} The log evidence as a function of the power-law slope using the new source pixelization scheme. \textbf{Bottom panel:} The log evidence as a function of the power-law slope using the source pixelization scheme of \citet{Nightingale2015}.}
    \label{fig:lklhd_curve}
\end{figure}

Finally, we assess the performance of our new source pixelization scheme by examining the smoothness of log evidence in response to small changes in the lens mass parameters. In the top panel of Fig.~\ref{fig:lklhd_curve}, we show the changes in log evidence with the slope of a power-law lens mass model for the mock lensed image, as presented in the right panel of Fig.~\ref{fig:voronoi_cells}. For the test, we employ a constant strength for the regularization, and we also constrain the solutions of the semi-linear inversion process to be non-negative. The minimum change in the power-law slope is set to be 0.0004. As shown, the log evidence changes smoothly with the power-law slope under the new source pixelization scheme, which incorporates natural neighbour interpolation and cross-like regularization. For comparison, we also plot the log evidence as a function of the slope under the old pixelization scheme in the bottom panel. As shown, the log evidence exhibits noisy behaviour with small changes in the power-law slope. At certain values, with a one-step change (0.0004) in slope, the corresponding change in log evidence can be as large as $\sim8$, which is statistically significant.  

\section{Normalized residuals of SLACS images when applying two other lens light models}

\begin{figure*}
    \includegraphics[width=2.0\columnwidth]{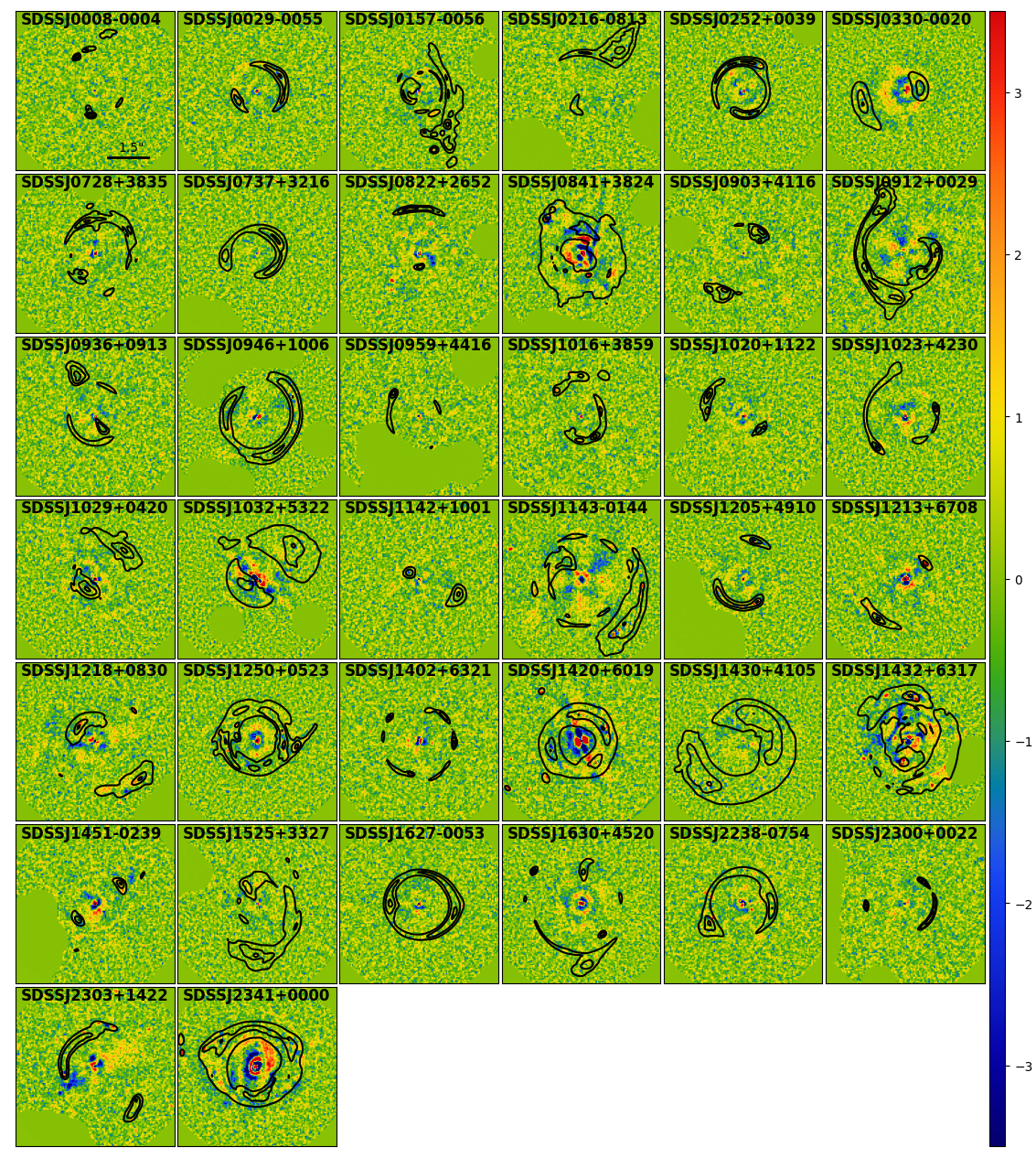}
    \caption{Best-fit normalized residuals for images of 38 SLACS lenses using the 3 Sersic lens light model. The colour bar range is from -3.5 to +3.5. The black contours are the same as those in Fig.~\ref{fig:slacs_results}.}
    \label{fig:slacs_results_3Sersics}
\end{figure*}

\begin{figure*}
    \includegraphics[width=2.0\columnwidth]{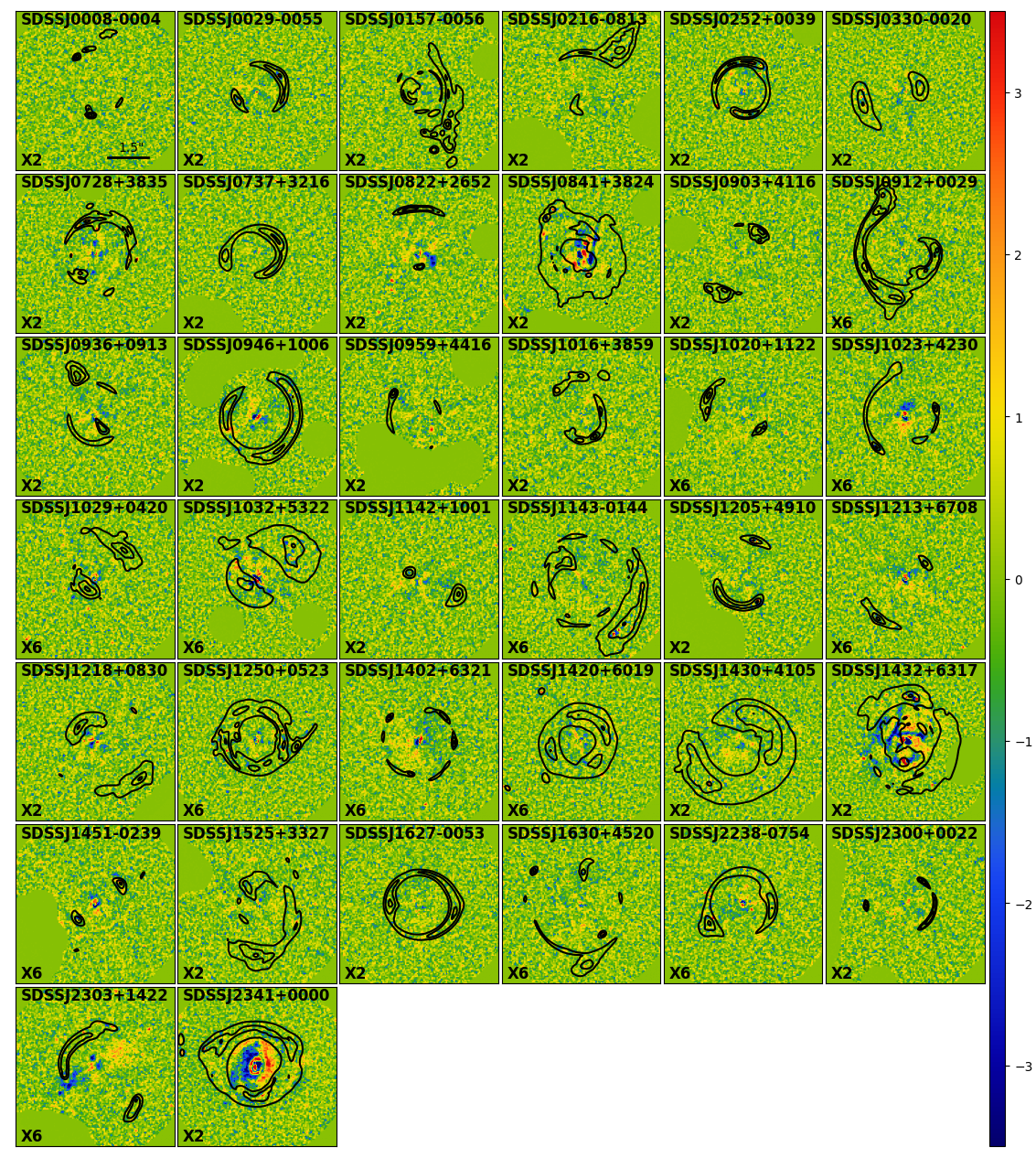}
    \caption{Best-fit normalized residuals for images of 38 SLACS lenses using the MGE lens light model without the ``point source'' set. The letter in the bottom left corner of every panel indicates the lens light model we used to fit the image. ``X2'' means a ``$2\times30$'' MGE lens light model while ``X6'' means a ``$6\times30$'' MGE lens light model. The colour bar range is from -3.5 to +3.5. The black contours are the same as those in Fig.~\ref{fig:slacs_results}.}
    \label{fig:slacs_results_MGE_nopoint}
\end{figure*}

\bsp	
\label{lastpage}
\end{document}